\documentclass{nature}

\usepackage{xcolor}
\usepackage{dcolumn}
\usepackage{bm}

\usepackage[
margin=0.5in,
]{geometry}

\usepackage{graphicx}
\usepackage{multicol}
\makeatletter
\let\saved@includegraphics\includegraphics
\AtBeginDocument{\let\includegraphics\saved@includegraphics}
\renewenvironment*{figure}{\@float{figure}}{\end@float}
\makeatother
\usepackage{subcaption}	
\usepackage{enumerate}
\usepackage[shortlabels]{enumitem}
\usepackage{amssymb}
\usepackage{amsthm}
\usepackage{amsmath,scalerel}

\usepackage{hyperref}
\hypersetup{colorlinks=true,linkcolor=blue,filecolor=blue,urlcolor=blue,citecolor=blue}
\usepackage[T1]{fontenc}
\usepackage[left]{lineno}

\usepackage{braket}
\renewcommand{\v}[1]{\ensuremath{\mathbf{#1}}} 
\let\baraccent=\= 
\renewcommand{\=}[1]{\stackrel{#1}{=}} 

\newcommand{\thmref}[1]{\hyperref[#1]{Theorem~\ref{#1}}}
\newcommand{\figref}[1]{\hyperref[#1]{Fig.~\ref{#1}}}
\newcommand{\figaref}[1]{\hyperref[#1]{Fig.~\ref{#1}a}}
\newcommand{\figbref}[1]{\hyperref[#1]{Fig.~\ref{#1}b}}
\newcommand{\figcref}[1]{\hyperref[#1]{Fig.~\ref{#1}c}}
\renewcommand{\eqref}[1]{\hyperref[#1]{Eq.~(\ref{#1})}}
\newcommand{\eqsref}[2]{\hyperref[#1]{Eq.~(\ref{#1})-(\ref{#2})}}
\newcommand{\appref}[1]{\hyperref[#1]{Appendix~(\ref{#1})}}

\makeatletter
\newcommand{\setword}[2]{%
	\phantomsection
	#1\def\@currentlabel{\unexpanded{#1}}\label{#2}%
}
\makeatother

\makeatletter
\def\@captype{figure}
\makeatother

\title{Quantum Locking of Intrinsic Spin Squeezed State in Earth-field-range Magnetometry}

\author{Peiyu Yang $^{1,2}$, Guzhi Bao $^{1,2,*}$, Jun Chen $^{1,2}$, Wei Du $^{1,2}$, Jinxian Guo $^{1,2}$, Weiping Zhang $^{1,2,3,4,\dagger}$}

\begin{document}
	
	\maketitle
	
	\begin{affiliations}
		\item {School of Physics and Astronomy, and Tsung-Dao Lee Institute, Shanghai Jiao Tong University, Shanghai 200240, China.}
		\item {Shanghai Branch, Hefei National Laboratory, Shanghai 201315, China.}
		\item {Shanghai Research Center for Quantum Sciences, Shanghai 201315, China.}
		\item {Collaborative Innovation Center of Extreme Optics, Shanxi University, Taiyuan, Shanxi 030006, China.}
		\item {* Corresponding author: guzhibao@sjtu.edu.cn}
		\item {$\dagger$ Corresponding author: wpz@sjtu.edu.cn}
	\end{affiliations}
	
	\begin{abstract}
		
		In the Earth-field range, the nonlinear Zeeman (NLZ) effect has been a bottleneck limiting the sensitivity and accuracy of atomic magnetometry from physical mechanism. To break this bottleneck, various techniques are introduced to suppress the NLZ effect. Here we revisit the spin dynamics in the Earth-field-range magnetometry  with the NLZ effect and identify the existence of the intrinsic spin squeezed state (SSS), generated from the coupling between nuclear and electron spins of each individual atom, with the oscillating squeezing degree and squeezing axis. Such oscillating features of the SSS prevent its direct observation and as well, accessibility to magnetic sensing. To exploit quantum advantage of the  intrinsic SSS in the Earth-field-range magnetometry, it's essential to lock the oscillating SSS to a persistent one. Hence we develop a quantum locking technique to achieve a persistent SSS, benefiting from which the sensitivity of the Earth-field-range magnetometer is quantum-enhanced. This work presents an innovative way turning the drawback of NLZ effect into the quantum advantage and opens a new access to quantum-enhanced magnetometry in the Earth-field range.
		
	\end{abstract}

	
	\begin{multicols}{2}
		[]
		
		\section*{Introduction}
		
		Sensitive measurements of magnetic fields in the Earth-field range is crucial to various real-world applications, including geological survey \cite{geo1,geo2,geo3,geo4,geo5}, biomedicine \cite{bio1,bio2,bio3,bio4,bio5}, fundamental physics experiments \cite{funda1,funda2,funda3,funda4,funda5}, and magnetic navigation \cite{navi1,navi2,navi3,navi4}. Alkali-metal atomic magnetometers  are the outstanding candidates for such missions because of their high sensitivity, reaching the fT/$\sqrt{\rm Hz}$  \cite{eq1,subfT,Fx,Earth1,Earth2,Earth3}. However, in the Earth-field range (50\,$\mu$T), the NLZ effect produces splittings and asymmetries of magnetic-resonance lines, leading to the signal reduction and heading errors, and brings a well-known bottleneck limiting the sensitivity and accuracy of atomic magnetometry \cite{spinlocking,dd,heading,all,2022,NLZ1,NLZ2,NLZ3}.
		
		In order to tackle this bottleneck, the intuitive thinking is to eliminate the NLZ effect, while maintaining the spin coherence. Such an operation renders an Earth-field-range magnetometer with a sensitivity at the standard quantum limit (SQL) \cite{spinlocking,dd,heading,all,2022,NLZ1,NLZ2,NLZ3}. Revisiting the spin dynamics in the Earth-field-range magnetometry with the NLZ effect, so far one has ignored an important fact that the intrinsic SSS exists due to the coupling of the electron spin and nuclear spin.
		However, as the atomic spin state sustains the continuous oscillation with the NLZ effect, the generated SSS exhibits the oscillating features on both the squeezing degree and squeezing axis, which prevent its direct observation and as well, accessibility to magnetic sensing.
		
		How to lock and utilize such an oscillating SSS becomes essential to turn the drawback into the advantage, opening a new access to the development of Earth-field-range quantum sensing.
		To lock the SSS to the axis of the measurement quantity with optimal squeezing degree, here we develop the quantum locking technique which incorporates a dynamic decoupling (DD) sequence optimized by machine learning. The main used differential evolution (DE) algorithm is a global optimization algorithm that was inspired by a natural selection strategy \cite{DE} and can easily be applied in numerical simulations or quantum experiments \cite{DEapp,DEapp2,DEapp3}. This model-free quantum locking technique has potential to be applied in other quantum technologies tackling decoherence such as performing error correction in quantum computation or maintaining the quantum state in quantum information. This intelligent Earth-field-range magnetometer, which automatically utilizes intrinsic SSS to achieve measurement sensitivity beyond the SQL, will inspires new ideas to those studying quantum metrology, Earth-field-range sensing, and quantum control, etc.
		
		\section*{Results}
		
		\subsection{Spin squeezing with the NLZ effect}
		
		The NLZ effect originates from the hyperfine interaction $\hat{i}\cdot\hat{j}$ and the interaction between atom and magnetic field, where $\hat{i}$ is the nuclear spin, $\hat{j}$ is the electron spin.
		The eigenvalues of atomic Zeeman sublevels in a magnetic field $B$ with electron quantum number $J=1/2$ are given by the Breit-Rabi formula \cite{BR,optically},
		\begin{equation}
			\begin{aligned}
				E_m=&-\frac{\Delta}{2(2I+1)}-g_I\mu_{B}mB\\
				&\pm \frac{\Delta}{2}(1+\frac{4m\xi}{2I+1}+\xi^2)^{1/2},
			\end{aligned}
			\label{BR}
		\end{equation}
		where $\xi=(g_{J}+g_{I})\mu_B B/\Delta$, $\Delta$ is the hyperfine-structure energy splitting, $I$ is the quantum number of nuclear spin, $g_J$ and $g_I$ are the electron and nuclear Landég factors, respectively, $\mu_B$ is the Bohr magneton, $B$ is the magnetic field intensity, and the $\pm$ refers to the $f=I\pm J$ hyperfine components for a single atom. In Earth's magnetic field, the nonlinear term in Eq.\,(\ref{BR}) cannot be neglectable, so we expand the eigenvalues to the second order of magnetic field $B$. For state $^{87}\text{Rb}5^{2}{S}_{1/2}f=2$, the energy level keeping only the magnetic field dependence is:
		\begin{equation}
			\begin{aligned}
				E_{m}\approx\frac{m\mu_BB}{2}-\frac{m^2\mu_{B}^{2}B^2}{4\Delta}.
			\end{aligned}
		\end{equation}
		Since the linear Zeeman effect shows linear dependence with magnetic quantum number $m$ and the NLZ effect shows quadratic dependence with $m$, the Hamiltonian of atom-magnetic field interaction with the leading magnetic field $B$ along the $\hat{\tilde{z}}$ axis can be described as:
		\begin{equation}
			\hat{H}_{B}\approx\hbar(\Omega_L\hat{f}_z-\omega \hat{f}_z^2),
			\label{E1}
		\end{equation}
		where $\hbar$ is Plank's constant, $\hat{f}_{z}$ is the angular momentum along $\hat{z}$ axis of a single atom, $\Omega_{L}=\gamma\,B=(\mu_{B}B)/(2\hbar)$ is the Larmor frequency, $\omega=(\mu_{B}B)^2/(4{\hbar^2}\Delta)$ is the quantum-beat revival frequency, $\gamma$ is the gyromagnetic ratio.
		The first term in Eq.\,(\ref{E1}) denotes the Larmor precession. The second term in Eq.\,(\ref{E1}) has the same form as the one-axis spin twisting introduced by Kitagaa and Ueda \cite{sss}. Note that the $\textrm{SSS}$ induced by the NLZ effect coming from the total angular momentum $\hat{f}$ of a single atom \cite{stark,AOC,Polzik2008}, whereas Kitagawa and Ueda refers to the collective angular momentum $\hat{F}$ of an ensemble of atoms.
		
		In the rotating frame, considering a coherent spin state (CSS) stretched along the $\hat{\tilde{y}}$ axis for $f=2$ system evolves under the free evolution Hamiltonian $\hat{\widetilde{H}}_{B}=-\hbar\omega\hat{\tilde{f}}_{z}^{2}$, where the operator $\hat{\tilde{o}}$ denotes the operator $\hat{o}$ in the rotating frame (Specific formulas are demonstrated in the Supplementary Note 2). For a spin state polarized perpendicular to the magnetic field, $\hat{\widetilde{H}}_{B}$ induces coupling between the eigenstates of the angular momentum along the direction of the polarized spin and causes squeezing of the spin state.
		The squeezing angle $\nu$, defined as the angle between the optimal squeezing axis and $\hat{\tilde{x}}$ axis, is calculated by $\nu =\pi/2- 1/2\arctan(Y/X) $, where $ X=1-\cos^{2f-2}(2\omega t) $, $ Y=4\sin(\omega t)\cos^{2f-2}(\omega t) $.
		The optimal squeezing axis of the SSS varies at different times.
		The green line in Fig.\,\ref{f1}(a) shows the minimum noise along the optimal squeezing axis at different times, illustrating that the squeezing degree changes over time as well.
		
		{\centering
			\includegraphics[width=0.5\textwidth]{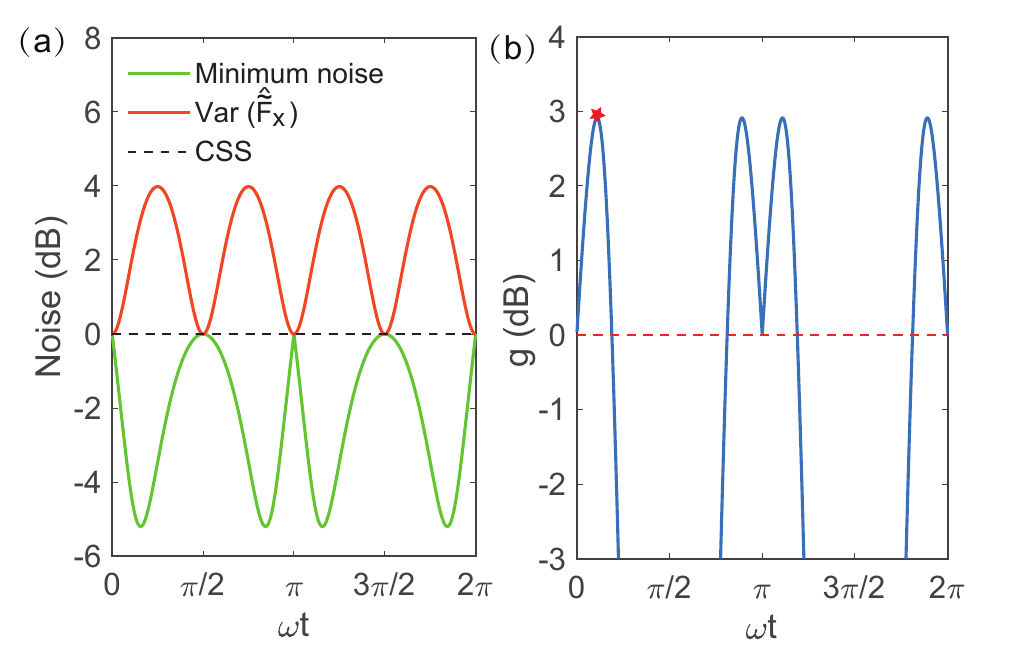}
			\caption{\textbf{The spin squeezing induced by the NLZ effect.} \textbf{(a)}~~The noise of the spin projections along the optimal squeezing axis (green line) and the noise of $\hat{\widetilde{F}}_x$ (red line) at different times. The dashed line is the noise of CSS $f/2$ for single atom. As the spin squeezing arises from the single-atom effect, the atom number $N$ is set to 1 in the simulation. \textbf{(b)}~~The metrological gain along the optimal squeezing axis at different times. The red star denotes optimal squeezing point to be locked.}
			\label{f1}}
		
		In an atomic magnetometer, the measurement quantity is the Larmor precession frequency, which is typically measured through finding the time between two zero crossings of a collective of oscillating spin component \cite{Fx}.
		For an atomic ensemble with $N$ atoms, the collective angular momentum components are $\hat{\widetilde{F}}_{x}=\sum\limits_{r=1}^{N}\hat{\tilde{f}}_{x}^{(r)}$ and $\hat{\widetilde{F}}_{y}=\textstyle\sum\limits_{r=1}^{N}\hat{\tilde{f}}_{y}^{(r)}$. While no atom-atom entanglement presents, the signal and noise of the corresponding magnetic measurement are scaled as $|\partial\langle\hat{\widetilde{F}}_x\rangle/\partial\theta_L|=\langle\hat{\widetilde{F}}_y\rangle=N\langle\hat{\tilde{f}}_y\rangle$ and $\Delta\hat{\widetilde{F}}_x=\sqrt{{\rm Var}(\hat{\widetilde{F}}_x)}=\sqrt{N{\rm Var}(\hat{\tilde{f}}_x)}$, respectively, where $\theta_L=\Omega_{L}T_{2}=\gamma BT_{2}$ is the phase of Larmor precession, $T_{2}$ is the coherence time of atomic spin.
		The sensitivity of the atomic magnetometry is
			\begin{equation}
				\begin{split}
					\Delta{B}&= \frac{\Delta\hat{\widetilde{F}}_x}{|\frac{\partial\langle\hat{\widetilde{F}}_x\rangle}{\partial\theta_L}||\frac{\partial\theta_L}{\partial{B}}|}=\frac{\Delta\hat{\widetilde{F}}_x}{\gamma{T}_{2}\langle\hat{\widetilde{F}}_y\rangle}=\frac{1}{\gamma{T}_{2}\sqrt{N}}\zeta,
					\label{sensitivity}
				\end{split}
			\end{equation}
			where $\zeta=\sqrt{{\rm Var}(\hat{\tilde{f}}_x)}/\langle\hat{\tilde{f}}_y\rangle$ is the spin-squeezing parameter of single atom. With CSS injection, Eq.(\ref{sensitivity}) determines the SQL for atomic magnetometry $\Delta{B}_{\rm SQL}= 1/(\gamma{T}_{2}\sqrt{2Nf})$ \cite{eq1,stark,Polzik2008,sss2,ASS,noise1,noise2} (see details in Supplementary Note 1).
			In this paper, we focus on the quantum enhancement of $\zeta$ which arises from nuclear-electronic spin entanglement \cite{stark,AOC,Polzik2008} (see details in Supplementary Note 3).
		
		To employ the SSS in quantum metrology, it is essential to match the angular momentum component with minimum noise to the measurement quantity. The red line in Fig.\,\ref{f1}(a) shows the noise ${\rm Var}(\hat{\widetilde{F}}_x)$ of the measurement quantity $\hat{\widetilde{F}}_x$ with NLZ effect, which always exceeds the fluctuation of CSS. That is because the optimal squeezing axis of this type of $\textrm{SSS}$ is not along the observable $\hat{\widetilde{F}}_x$.
		We can rotate the SSS along the -$\hat{\tilde{y}}$ axis by $\nu$ to align the optimal squeezing axis with the axis of $\hat{\widetilde{F}}_x$, thereby ensuring that $\hat{\widetilde{F}}_x$ has the minimum noise. The quantum enhancement of magnetic field measurement arises from squeezing of the noise.
		
		After the rotation operation with angle $\nu$, we calculate the resulting single-shot sensitivity of the magnetic field measurement with $N$ atoms \cite{beatingSQL}
		\begin{equation}
				\begin{split}
					\Delta{B}_{\rm SSS}
					=&\frac{\sqrt{{\rm Var}(\hat{\tilde{f}}_{x,\nu})}}{\gamma{T}_{2}\sqrt{N}\langle\hat{\tilde{f}}_{y}\rangle}\\
					=&\frac{\sqrt{1+\frac{1}{2}(f-\frac{1}{2})(X-\sqrt{X^2+Y^2})}}{\gamma{T}_{2}\sqrt{N}\sqrt{2f}(\cos\omega t)^{2f-1}},
					\label{sp}
				\end{split}
			\end{equation}
			see details in Supplementary Note 2.
		In order to intuitively represent the advantage of the SSS injection, we use the metrological gain defined as $g=(\Delta{B}_{\rm SQL}/\Delta{B}_{\rm SSS})^2$ \cite{gain} to represent the quantum enhancement for the sensitivity.
		The metrological gain is then $g=(\cos\omega t)^{2(2f-1)}/[1+(1/2)(f-1/2)(X-\sqrt{X^2+Y^2})]$.
		Since the spin squeezing arises from the single-atom effect, the metrological gain is independent of $N$. For $ f=2 $ system, the metrological gain is shown in Fig.\,\ref{f1}(b). The maximum metrological gain defined as the optimal squeezing point approaches 2.9\,dB at $\omega t =0.34$. Note that the maximum gain [Fig.\,\ref{f1}(b)] doesn't occur at the minimum noise [Fig.\,\ref{f1}(a)], because the signal and the noise have different time dependencies.
		
		{\centering
			\includegraphics[width=0.48\textwidth]{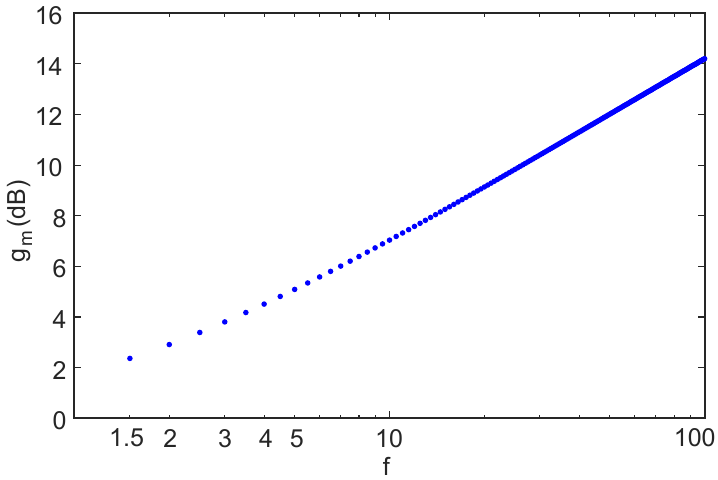}
			\caption{\textbf{The scalability of the spin squeezing.} The maximum metrological gain $g_m$ varies with $f$.}
			\label{f2}}
		The maximum metrological gain $g_m$ with respect to $\omega t$ is scalable with the quantum number of total angular momentum $f$ for a single atom. Figure\,\ref{f2} plots $g_m$ as a function of $f$. Relatively large values of $f$ available are $f$ = 4 in the ground state of Cs and $f$ = 12.5 in a metastable state of Dy \cite{Dy}, corresponding to a maximum metrological gain  by\,4.5 dB and 7.7\,dB, respectively, which are comparable to the reported ones with multi-atom collective spin squeezing \cite{sss2,BEC3}. The maximum metrological gain can scale up even higher with larger $f$ such as in Rydberg atoms and in molecules \cite{stark}.

		\subsection{Quantum spin locking}
		
		The $\textrm{SSS}$ produced by the NLZ effect could help improve the sensitivity of atomic magnetometry. However, because the NLZ effect always exists, the SSS keeps evolving. To lock the SSS at the optimal squeezing point [the star in Fig.\,\ref{f1}(b)], we develop the quantum locking method, which is a machine-learning-assisted DD technique.
		
		We set the wavefunction of the SSS to be locked as $|\Psi(0)\rangle$. Figure \ref{f3} shows the principle of the quantum spin locking scheme in the form of the real-time fidelity $\mathcal{F}_{r} ={\rm Tr}|\sqrt{\sqrt{\rho_{0}}\rho_{t}\sqrt{\rho_{0}}}|^2$ between the SSS and the state at time $t$. $\rho_0=|\Psi(0)\rangle\langle\Psi(0)|$ is the density matrix of the SSS. $\rho_{t}$ is the density matrix of spin state at time $t$. If the SSS evolves freely with the NLZ effect, the SSS collapses and revives periodically, as the blue line in Fig.\,\ref{f3} shown.
		While, in the quantum locking scheme, utilizing the periodic property of spin evolution under the NLZ effect, the SSS can be locked by taking the shortcut from $\Psi(t_1)$ to $\Psi(2\pi-t_1)$, which can be achieved by the spin evolution with the DD sequence $\hat{\widetilde{U}}_{DD}(T)$ (as the red line in Fig.\,\ref{f3} shown), where $T$ is the period of one DD cycle.
		More detail is discussed in Supplementary Note 8.
		
		{\centering
			\includegraphics[width=0.48\textwidth]{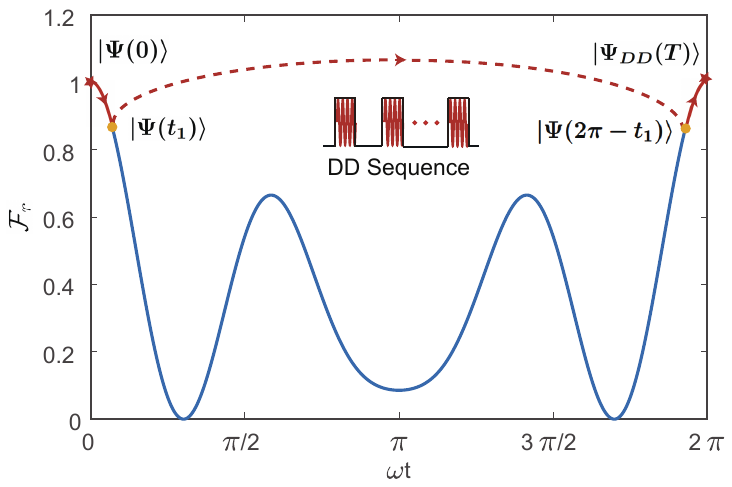}
			\caption{\textbf{The principle of the quantum spin locking scheme.} The red star at the starting point is the SSS $|\Psi(0)\rangle$ to be locked. The blue line represents the real-time fidelity of the SSS oscillating at the period $2\pi$ with the NLZ effect. The red line is the path of the state evolution in the DD process. The SSS firstly evolves to $|\Psi(t_1)\rangle$, then slips to $|\Psi(2\pi-t_1)\rangle$ and finally revivals to the state $|\Psi_{DD}(T)\rangle$. $T$ is the period of the DD sequence and is set substantially shorter than the revival period $2\pi/\omega$. $\Omega_L=2\pi$,  $\Omega_{L}/\omega=20000$, $T\Omega_{L}=2\pi\times2000$.}
			\label{f3}}
		
		The DD sequence consists $p$ instaneous rotation pulses $\hat{\Pi}^{d}(\chi_{i})\,(i=1...p)$ following the free evolution $\hat{\widetilde{U}}_F(T/p)$, where $\hat{\Pi}^{d}(\chi_{i})=e^{-i\chi_{i}\hat{\tilde{f}}_{d}}$ is the pulsed rotation evolution operator,
		$\hat{\widetilde{U}}_F(T/p)=e^{-i\hat{\widetilde{H}}_{B}T/p\hbar}$ is the free evolution operator. The pulse is applied by a radio-frequency (RF) magnetic field in the laboratory frame (a static magnetic field in the rotating frame).
		$d = \pm \hat{\tilde{x}}, \pm \hat{\tilde{y}}, \pm \hat{\tilde{z}}$ is the rotation direction in the rotating frame which is decided by the phase of RF field, $\chi_{i}$ is the rotation angle of the $i$-th pulse. To ensure that the atomic spin remains stretched along the $\hat{\tilde{y}}$ direction and that as many atomic spins as possible can be used to sense the magnetic field, the pulses in the DD sequence are all along the -$\hat{\tilde{y}}$ axis. Hence, the DD sequence takes the form
		\begin{equation}
			\begin{aligned}
				\hat{\widetilde{U}}_{DD}(T) =& \hat{\Pi}^{-\hat{\tilde{y}}}(\chi_{p})\hat{\widetilde{U}}_F(\frac{T}{p})\hat{\Pi}^{-\hat{\tilde{y}}}(\chi_{p-1})\hat{\widetilde{U}}_F(\frac{T}{p})\\
				&\cdots\hat{\Pi}^{-\hat{\tilde{y}}}(\chi_{2})\hat{\widetilde{U}}_F(\frac{T}{p})\hat{\Pi}^{-\hat{\tilde{y}}}(\chi_{1})\hat{\widetilde{U}}_F(\frac{T}{p}).
				\label{dd1}
			\end{aligned}
		\end{equation}
		
		{\centering \includegraphics[width=0.48\textwidth]{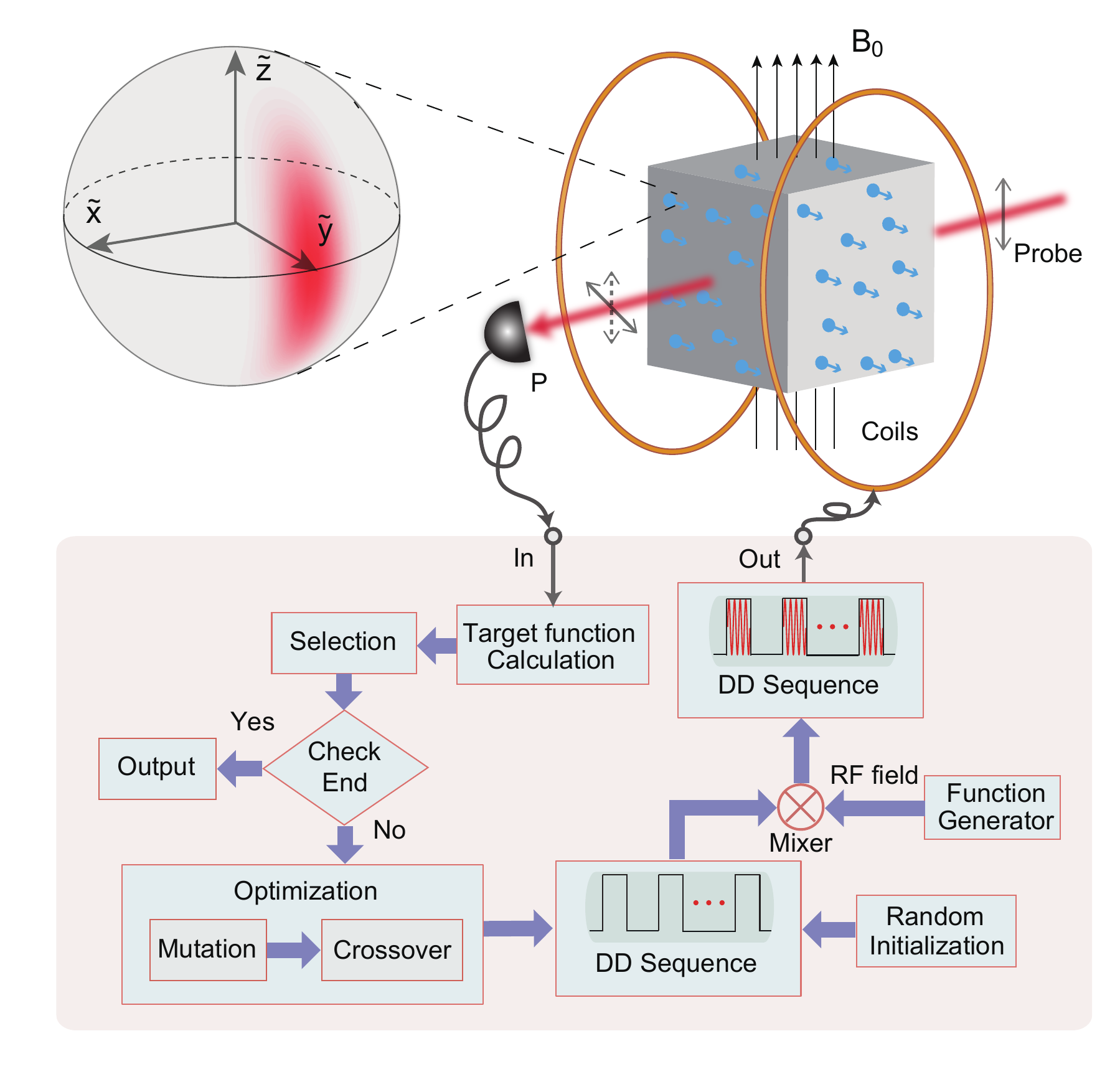}
			\caption{\textbf{System of the quantum locking scheme.} With the leading magnetic field $B_0$ along the $\hat{\tilde{z}}$ axis, the atomic spins oscillate in the $\hat{x}-\hat{y}$ plane at Larmor frequency in the laboratory frame and are orientated along the $\hat{\tilde{y}}$ axis in the rotating frame. With the NLZ effect, the spins are squeezed along the optimal squeezing axis. After the rotation operation with angle $\nu$, the $\hat{\widetilde{F}}_{x}$ is squeezed as represented on the Bloch sphere. A sequence of pulses optimized by the DE algorithm are applied in the form of the RF field pulses along the -$\hat{\tilde{y}}$ axis to lock the SSS. P is the polarimeter used to measure the spin projection of the atomic state for the target function calculation. The end condition is when we reach the maximum generation. More detail is discussed in Methods.}
			\label{f4}}
		
		For a specific angular momentum $f$, the DD performance is a function of parameters $T$, $p$, and $\chi_i\,(i=1...p)$.
		In the experiment, there are some restrictive conditions for the values of these parameters.
		The DD period $T$ should be substantially shorter than the revival period $2\pi/\omega$ to ensure that the SSS experiences enough DD cycles and keeps squeezed most of the time.
		Theoretically, the rotation angle $\chi_i$ is arbitrary. Considering the running speed of the algorithm and the convenience of pulse design in the experiment, $\chi_i$ is set to be $n\pi/36$ (step length is 5 degrees), where $n$ is an integer.
		The relationship between the rotation angle $\chi_{i}$ and the pulse length $t_{i}$, is formulated as $\chi_{i}=\int_{0}^{t_i}\Omega_i(t)dt$, where $\Omega_i(t)$ denotes the energy distribution of the $i$th pulse. The pulse length $t_i$ should be considerably shorter than the free evolution time $T/p$ to satisfy the instaneous rotation condition and substantially longer than the Larmor period to satisfy the rotation wave approximation condition \cite{RWA}. Considering the Larmor frequency, the quantum-beat revival frequency of $^{87}$Rb in the Earth-field range and the restrictive conditions mentioned above, we choose $\Omega_{L}/\omega=20000$, $T\Omega_{L}=2\pi\times2000$, and $p=10$ for the DD sequence optimization.
		
		We introduce the final state $\mathcal{F}_{f}={\rm Tr}|\sqrt{\sqrt{\rho_{0}}\rho_{T}\sqrt{\rho_{0}}}|^2$ to evaluate the DD performance \cite{F1,F2,F3}, where $\rho_{T}=\hat{\widetilde{U}}_{DD}(T)|\Psi(0)\rangle\langle\Psi(0)|\hat{\widetilde{U}}^{\dagger}_{DD}(T)$ is the density matrix of spin state after one DD sequence. For the locking of the SSS, our goal is to determine the optimal $\chi_i$ in the DD sequence so that the fidelity $\mathcal{F}_{f}$ is as close to 1 as possible.
		The DE algorithm is used to obtain the DD sequence. The target function is the infidelity $1-\mathcal{F}_{f}$. The spin state is measured by the Faraday interaction between the probe light and the atomic system (see Methods for details). The fidelity can be obtained by the state tomography through varying the propagation direction of the probe light \cite{Tomography1,Tomography2,Tomography3}.
		Our pulse sequence design task can be seen as a kind of multi-parameter optimization with discrete variables. As a gradient-free algorithm, the DE algorithm is a powerful choice, which is not easily trapped in a local optimum solution \cite{DE}. The framework of the proposed approach is shown in Fig.\,\ref{f4}. For given $T$ and $p$, we generate an initial population containing a series of DD sequences with random rotation angles. The DD sequences are applied to the atomic system by the coils in the form of the RF field pulses. The main operations are mutation, crossover and selection. In the mutation operation, three DD sequences are randomly selected from the population as the mutation sources and combined to reproduce the mutation sequence [see Eq.\,(\ref{mutation}) in Methods]. The rotation angles in the original DD sequence are partially replaced by the ones in the mutation sequence to derive the test sequence [see Eq.\,(\ref{crossover}) in Methods], which is the crossover operation. In the selection process, the target function values of the test sequence and the original sequence are compared, and the sequence with the better value is retained to the next optimization loop. The three operations mentioned above are performed once on each DD sequence in every loop to update the population. After a number of iterations, the algorithm gradually converges to the optimal sequence
		\begin{equation}
			\begin{aligned}
				\hat{\widetilde{U}}_{DD}(T)& = \hat{\Pi}^{-\hat{\tilde{y}}}(\frac{\pi}{36})\hat{\widetilde{U}}_F(\frac{T}{10})\hat{\Pi}^{-\hat{\tilde{y}}}(\frac{4\pi}{9})\hat{\widetilde{U}}_F(\frac{T}{10})\\
				&\times\hat{\Pi}^{-\hat{\tilde{y}}}(\frac{\pi}{4})\hat{\widetilde{U}}_F(\frac{T}{10})\hat{\Pi}^{-\hat{\tilde{y}}}(\frac{4\pi}{9})\hat{\widetilde{U}}_F(\frac{T}{5})\\
				&\times\hat{\Pi}^{-\hat{\tilde{y}}}(\frac{\pi}{36})\hat{\widetilde{U}}_F(\frac{2T}{5})\hat{\Pi}^{-\hat{\tilde{y}}}(\frac{\pi}{6})\hat{\widetilde{U}}_F(\frac{T}{10}),
				\label{dd}
			\end{aligned}
		\end{equation}
		in which $T$ is set as 1/10 revival periods.
		
		The dynamic evolution of the atomic state is illustrated by calculating the quasi-probability distribution \cite{sss}. Figure \ref{f5}(a)-(c) represent the generation of the SSS and (c)-(o) describe the evolution of the SSS during the DD sequence in Eq.(\ref{dd}). The fidelity between the states (c) and (o) is 99.99\%, which proves the success of this DD sequence.
		
		{\centering
			\includegraphics[width=0.48\textwidth]{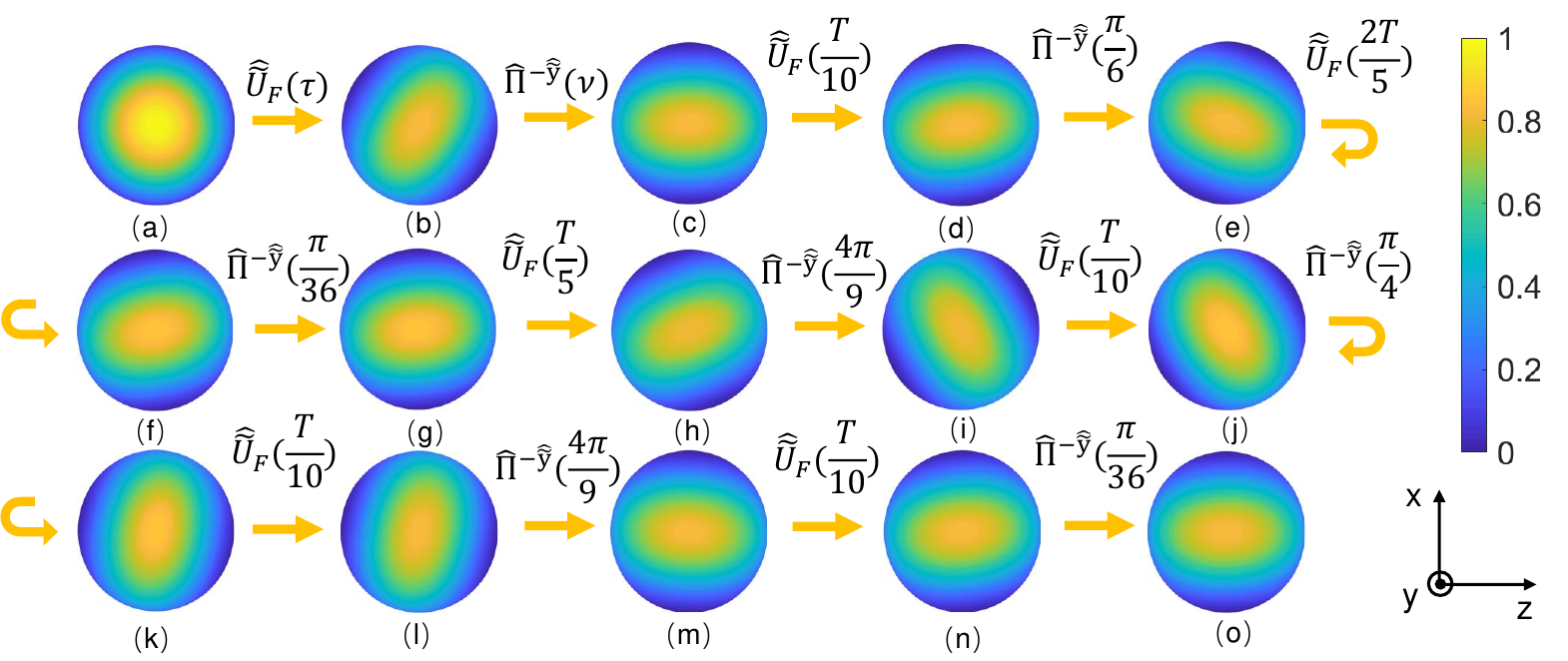}
			\caption{\textbf{Evolution of the atomic spin state.} The quasi-probability distribution of the atomic spin state $Q(\theta,\phi) = |\langle\theta,\phi|\Psi (t)\rangle|^2$, where $ |\theta,\phi\rangle $ is the CSS, $ \theta $ is the polar angle ranging from $ [0,\pi] $, $ \phi $ is the azimuthal angle ranging from $ [0,2\pi] $, and $ |\Psi (t)\rangle $ is the atomic wavefunction at time $ t $. \textbf{(a)}~~The initial CSS stretched along the $\hat{\tilde{y}}$ axis. \textbf{(b)}~~The optimal SSS. \textbf{(c)}~~The SSS with the optimal squeezing axis along the $\hat{\tilde{x}}$ axis after the rotation operation. \textbf{(d)}-\textbf{(o)}~~The evolution of the SSS (c) during one DD cycle. $\tau$ is the time when the most squeezed spin state occurs and $\omega \tau=0.34$. T is the period of the DD sequence.}
			\label{f5}}
		
		\subsection{Magnetic field measurement with the SSS}
		To measure the magnetic field, a linearly polarized probe light is utilized to measure the spin component projection along the $\hat{x}$ axis by the first order atom-light interaction (see Methods for details).
		After the probe light propagating through the atomic sensors, the polarization angle of the probe light is rotated. From the polarization rotation signal detected by a polarimeter, we therefore obtain the spin component $\hat{F}_{x}$ which oscillates at Larmor frequency of leading magnetic field, as shown in Fig.\,\ref{f4}.
		Here, we measure the magnetic field perturbation $\delta{B}$ in the leading magnetic field $B_{0}$. In the locking stage, the frequency of the RF field is the same as the Larmor frequency of $B_{0}$. In the measurement stage, without the RF field, the Larmor precession frequency can directly reflect $\delta{B}$.
		
		There are two protocols to apply the probe light pulse. The SSS can be locked during an interval of darkness permitting the spin state to precess in the presence of the static magnetic field and DD pulses. A probe pulse is then applied for readout \cite{OL}, as shown in the protocol I in the Fig.\,\ref{f6}(a). In this condition, the power broadening and back-action induced by the probe light is avoided by sacrificing the measurement time $t_{p1}$. While, in a conventional magnetometry, the probe light is applied in the whole time $t_{p2}$ and monitor the Larmor procession, as shown in the protocol II in the Fig.\,\ref{f6}(a). In this protocol, high fidelity during the whole measurement process is required and the optimal DD sequence is obtained by maximize the $\overline{\mathcal{F}_{r}}$, where $\overline{\mathcal{F}_{r}}$ is the average real-time fidelity during the DD operation.
		
		{\centering
			\includegraphics[width=0.48\textwidth]{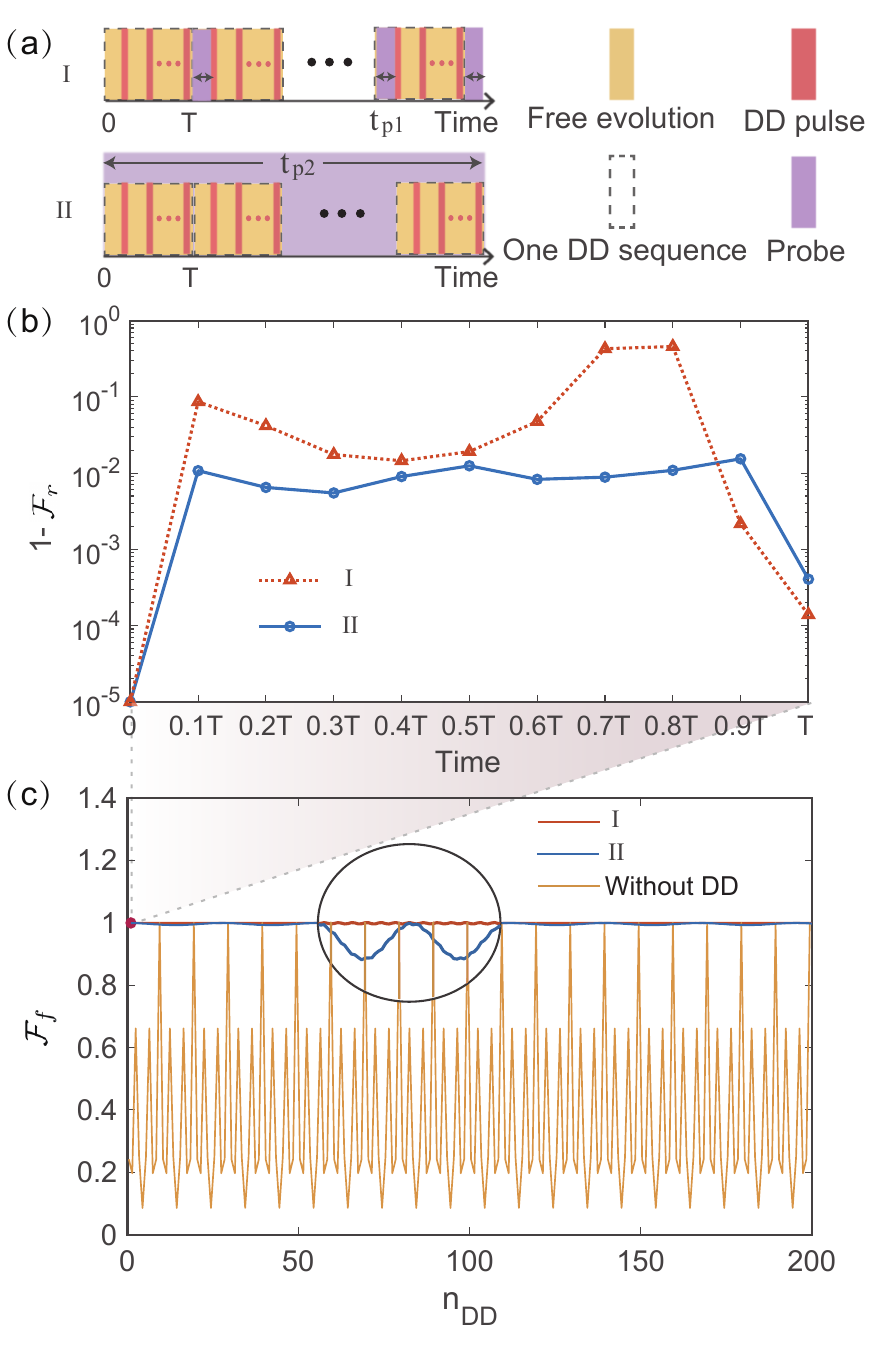}
			\caption{\textbf{Quantum locking for different measurement protocols.} \textbf{(a)}~~The sequence in different measurement protocols. $t_{p1}$ and $t_{p2}$ are the total time of the probe light in different protocols. \textbf{(b)}~~The real-time infidelity $1-\mathcal{F}_{r}$ after each rotation operation in one DD sequence in different measurement protocols. T is set as 1/10 revival periods. \textbf{(c)}~~The final fidelity $\mathcal{F}_f$ during 200 DD cycles in different protocols. Inset is the zoom in picture. $n_{DD}$ denotes the number of the DD cycles.}
			\label{f6}}
		
		Figure \ref{f6}(b) compares the real-time infidelity $1-\mathcal{F}_{r}$ after each rotation operation in one DD sequence for the two measurement protocols. In protocol I, the $\overline{\mathcal{F}_{r}}$ during the DD operation is low, but the spin state is well locked at the end of the DD sequence with a final fidelity of $99.99\%$. In protocol II, the fidelity of the final state $99.83\%$ is lower than that in protocol I, while the $\overline{\mathcal{F}_{r}}$ is better.
		
		An effective DD sequence is also essential to ensure a high fidelity for a long-time duration, which guarantees sufficient time to perform the measurement.
		Without the DD sequence, the fidelity periodically oscillates due to the NLZ effect as shown in Fig.\,\ref{f6}(c); while with DD sequence, the fidelity remains in the range $[99.97\%,99.99\%]$ in protocol I and $[99.28\%,99.99\%]$ in protocol II. The long-time stability of the locked SSS in protocol I is better than that in protocol II, even during several hundred cycles. The quantum spin locking scheme is flexibly adapted to different measurement protocols as we choose appropriate target function depending on the requirements for the SSS in the measurement process.
		
		In addition, a magnetic field gradient is inevitable in a natural environment. The phase perturbation due to an inhomogeneous magnetic field accumulates over time and causes a small spin-orientation broadening along the spin state. Dephasing due to an inhomogeneous magnetic field can be largely suppressed by using the Hahn echo \cite{dd,Hahn}, where a $\hat{\Pi}^{\tilde{y}}(\pi)$ pulse is applied during the free-evolution interval. In quantum locking scheme, to remove the magnetic field gradient simultaneously, the $\hat{\Pi}^{\tilde{y}}(\pi)$ pulse could be added in each free evolution $\hat{\widetilde{U}}_F(T/p)$ of DD sequence. Notably, the $\hat{\Pi}^{\tilde{y}}(\pi)$ pulses have no influence on the quantum locking scheme \cite{dd}.
		
		\section*{Discussion}
		
		\noindent To conclude, in this paper,
		the intelligent Earth-field-range magnetometer, which automatically utilizes intrinsic SSS to achieve measurement sensitivity beyond the SQL is proposed.
		We turn the NLZ effect which limits the sensitivity of the Earth-field-range magnetometer into a valuable technique for quantum enhancement of magnetic sensing. Due to the continuous collapse and revival of the atomic state induced by the NLZ effect, the SSS oscillates over time and is not typically used in experiments. This issue can be solved by locking the SSS with periodic pulsed modulations. The pulse sequence is calculated by the DE algorithm.
		
		There are three steps to build such a quantum-enhanced magnetometer in the Earth-field range. First, the state evolves freely to generate the SSS; then, at the time of the optimal squeezing point, a rotation operation is performed to align the optimal squeezing axis with the measurement axis; and finally, a DD sequence is applied to lock the SSS and the magnetic measurement is implemented. With the help of the SSS injection, the noise of the magnetic resonance can be reduced; in addition, the application of the quantum locking technique cancels the extra NLZ effect, which increases the magnetic resonance's amplitude and narrows its linewidth. Supposing the atoms in a 1\,${\rm cm^3}$ cell working in the room temperature, the atom number is $1.42\times10^{10}$ \cite{density}. With the coherence time of 10\,ms, the SQL is 5.99\,fT/$\sqrt{\text{Hz}}$. For $^{87}$Rb in the Earth's magnetic field ($50\mu$T), due to the NLZ effect, the spin state undergoes rapid collapse with an effective coherence time of $1.82$\,ms, which leads to a sensitivity of 14.04\,fT/$\sqrt{\text{Hz}}$. By using the quantum locking technique, with the benefit of both maintaining the coherence time and spin squeezing, the sensitivity reaches 4.29\,fT/$\sqrt{\text{Hz}}$. Such quantum-enhanced magnetometry is promising for practical applications in the Earth-field range.
		
		The quantum locking technique proposed in this paper can, without loss of generality, be extended to lock the other spin system with quadratic interaction \cite{quad1,quad2} or large-$f$ SSS with high metrological gain.
		In this work, $B_{0}$ is fixed and $\delta{B}$ is considered as a minor perturbation.
		In the case of substantial variation of the magnetic field, the frequency of the RF field applied each time needs to be adjusted based on the result of the previous measurement.
		The measured Larmor frequency serves as feedback to correct the RF field frequency with the Larmor frequency. Similar closed-loop schemes have been achieved in conventional optically pumped magnetometry by self-oscillating \cite{selfos} or phase lock loop \cite{phaseloop} which lock the frequency of driving field to the Larmor frequency. Then the uncertainty of the previous measurement would influence the overall measurement performance. In our system, the single-shot sensitivity is at the level of 1\,pT. This level of uncertainty does not notably impede the metrological gain and DD operations. More detail is discussed in Methods.
		Here we study the SSS generated from a single atom. By generating entanglement between atoms, the metrological gain can be further improved \cite{BEC3}. The protocol proposed here works as a scalar magnetometer by measuring the Larmor frequency of atomic spins which is intrinsically sensitive to the magnitude of an applied field rather than its projection along a particular direction. There are several ways to derive information of magnetic field direction, converting a scalar magnetometer into a vector magnetometer \cite{Vector1, Vector2}.
		With the magnetic field much larger than Earth-field range, higher order nonlinear effect would arise. The spin dynamic in this condition needs to be further studied.
		
		\begin{methods}
			\subsection{Optimization algorithm}
			In our work, a DD sequence is used to lock the SSS with high fidelity. The DE algorithm is used to determine the optimal rotation angles in the DD sequence. The target function $f(\vec{x})=1-\mathcal{F}_{f}$ is served as the cost function of the DE algorithm, where $\mathcal{F}_{f}$ is the fidelity calculated after the DD sequence. The variable $\vec{x}$ denotes a DD sequence containing $p$ rotation angles and is written in the form of a vector as
			\begin{equation}
				\vec{x}=(x^{1},x^{2},...,x^{p}),
			\end{equation}
			where $x^{i} (i = 1,2,...,p)$ is the $i$-th rotation angle in the sequence.
			
			The domain of the function $f(\vec{x})$ with maximum rotation angle $\vec{x}_{max}$ and the minimum rotation angle $\vec{x}_{min}$ is defined as
			\begin{equation}
				\left\{
				\begin{aligned}
					\vec{x}_{max} &=(x^{1}_{max},x^{2}_{max},...,x^{p}_{max}), \\
					\vec{x}_{min} &=(x^{1}_{min},x^{2}_{min},...,x^{p}_{min}).
				\end{aligned}
				\right.
			\end{equation}
			Our target is to find the minimum of the target function $f(\vec{x})$ within the defined domain.
			
			First, an initial population containing $M$ DD sequences with $p$ random rotation angles is generated. The $j$-th sequence in the $k$-th loop of the population can be expressed as
			\begin{equation}
				\begin{aligned}
					&\vec{x}_{j,k}=(x^{1}_{j,k},x^{2}_{j,k},...,x^{p}_{j,k}), \\
					&j=1,2,...,M,\\
					&k=1,2,...,Q,
				\end{aligned}
			\end{equation}
			where $Q$ is the maximum number of the loop.
			The $j$-th sequence in the first loop $\vec{x}_{j,1}$ is obtained randomly:
			\begin{equation}
				x^{i}_{j,1}=x^{i}_{min}+rand(0,1) \cdot (x^{i}_{max}-x^{i}_{min}),
			\end{equation}
			where $rand(0,1)$ is a uniform random number between 0 and 1.
			
			The next operation is the mutation operation. Three DD
			sequences are randomly selected from the population as
			the mutation sources and combined to reproduce the mutation sequence. The mutation sequence $\vec{v}_{j,k}$ for the $j$-th sequence in the $k$-th loop is obtained by
			\begin{equation}
				\vec{v}_{j,k}=\vec{x}_{r^{j}_{1},k}+F_{s} \cdot (\vec{x}_{r^{j}_{2},k}-\vec{x}_{r^{j}_{3},k}),
				\label{mutation}
			\end{equation}
			where $r^{j}_{1}$, $r^{j}_{2}$ and $r^{j}_{3}$ are randomly generated integers that are not equal to $j$, $F_{s}$ is a preset scaling factor in the range $0<F_{s}<1$.
			
			Then, the crossover operation is applied. The rotation angles in the original DD sequence $\vec{x}_{j,k}$ are partially replaced by the ones in the mutation sequence $\vec{v}_{j,k}$ to derive the test sequence $\vec{u}_{j,k}$ as follows:
			\begin{equation}
				u^{i}_{j,k}=\left\{
				\begin{array}{rcl}
					v^{i}_{j,k},       &      & {\rm if}\,(i = i_{rand})\,\,{\rm or} \,\,(r^{i}_{j} < P_{cr}),\\
					x^{i}_{j,k},    &      & {\rm otherwise},
				\end{array} \right.
				\label{crossover}
			\end{equation}
			where $i_{rand}$ is a random integer, $r^i_j$ is a random number between 0 and 1, and $P_{cr}$ is the predefined crossover probability.
			
			Finally, in the selection process, the target function
			value $f(\vec{u}_{j,k})$ is compared with $f(\vec{x}_{j,k})$. The sequence with smaller value is retained in the next loop i.e.
			\begin{equation}
				\vec{x}_{j,k+1}=\left\{
				\begin{array}{rcl}
					\vec{u}_{j,k},       &      & f(\vec{u}_{j,k}) < f(\vec{x}_{j,k}),\\
					\vec{x}_{j,k},    &      & {\rm otherwise}.
				\end{array} \right.
			\end{equation}
			
			The three operations are performed once on each DD sequence in every
			loop to update the population. After a number of iterations, the algorithm
			gradually converges to the optimal sequence until the number of loop reaches a preset threshold.
			
			In our case,  the dimension $p$ of $\vec{x}$ is set to 10 and the rotation angle is set to be $n\pi/36$, where $n$ is an integer. Considering the running speed of the algorithm, the domain of $\vec{x}$ is set to $[0,\pi/2]$. The size of the population $M$ is set to 100. The maximum number of loop $Q$ is set to 1000. The scaling factor $F_{s}$ is set to $0.5$. The crossover rate $P_{cr}$ is set to $0.3$.
			
			\subsection{Spin state measurement}
			To measure the atomic spin state and thus obtain the Larmor frequency of the atomic spin, a far-off resonant and linear-polarized light propagating along the $\hat{x}$ axis is introduced. The proprieties of the probe light can be described by the Stokes operators $\hat{S}_{x}=1/2(\hat{a}_{x}^{\dagger }\hat{a}_{x}-\hat{a}_{y}^{\dagger }\hat{a}_{y})$, $\hat{S}_{y}=(\hat{a}_{x}^{\dagger }\hat{a}_{y}+\hat{a}_{y}^{\dagger }\hat{a}_{x})$, $\hat{S}_{z}=i/2(\hat{a}_{y}^{\dagger }\hat{a}_{x}-\hat{a}_{x}^{\dagger }\hat{a}_{y})$.
			The Hamiltonian of atom-light interaction $\hat{H_{L}}$ can be approximated to the first order:
			\begin{equation}
				\hat{H}_{L}=\kappa\hat{S}_{z}\hat{F}_{x},
			\end{equation}
			where, $\kappa$ is the first order coupling constant of the atom-light interaction. The input-output relation of $\hat{S}_{y}$ is:
			\begin{equation}
				\hat{S}^{out}_{y}=\hat{S}^{in}_{y}+\kappa\hat{S}_{x}\hat{F}_{x}.
			\end{equation}
			With this interaction, the spin projection along the $\hat{x}$ axis can be measured by monitoring the $\hat{S}^{out}_{y}$ of the probe light. The other spin components can be measured by changing the propagation direction of the probe light. The fidelity can be derived by the state tomography.
			
			We also notice that the atom-light interaction would induce back action to the atomic spins. The input-output relation of spin operators is:
			\begin{subequations}
				\begin{align}
					\frac{\partial}{\partial\,t}\hat{F}_{x}&=0,\\
					\frac{\partial}{\partial\,t}\hat{F}_{y}&=-\kappa\hat{S}_{z}\hat{F}_{z},\\
					\frac{\partial}{\partial\,t}\hat{F}_{z}&=\kappa\hat{S}_{z}\hat{F}_{y}.
				\end{align}
			\end{subequations}
			Although the probe light is linearly polarized i.e. $\langle\hat{S}_{z}\rangle=0$, it will have quantum fluctuations in the degree of the circular polarization. With a macroscopic spin polarized along the $\hat{z}$ ($\hat{y}$) axis, the fluctuation of optical field is mapped to the spin fluctuation $\hat{F_{y}}$ ($\hat{F_{z}}$). Since the spin in our system is oscillating on the $\hat{x}-\hat{y}$ plane, the back action is always acted on $\hat{F}_{z}$ which is the anti-squeezed component of the SSS mentioned in the last section. So the back action does not affect the squeezing of the spin state.
			
			\subsection{Closed-loop scheme with substantial variation of the magnetic field}
			In the magnetic field measurement, we measure the magnetic field perturbation $\delta{B}$ in the leading magnetic field $B_{0}$. In the main text, $B_{0}$ is fixed and $\delta{B}$ is considered as a minor perturbation. In this case, the frequency of the RF field is set to the same as the Larmor frequency of $B_{0}$ and the angle $\nu$ is also derived from $B_{0}$.
			While, in the case of substantial variation of the magnetic field, if the RF field frequency and angle $\nu$ are still derived from $B_0$, the initially generated SSS and the DD performance will be affected by the variation $\delta{B}$ of the magnetic field. This issue can be addressed using closed-loop scheme. The measured Larmor frequency is used as feedback to adjust the frequency of RF field in order to lock it to the Larmor frequency, as shown in Fig\,\ref{AP}.
			Then the uncertainty of previous measurement would influence the measurement performance.
			In each measurement process, despite possible degradation caused by $\delta B$ in the SSS generation and the DD performance, there exists a range of $\delta B$ for maintaining the metrological gain. To guarantee quantum enhancement in each measurement, the single-shot measurement uncertainty should be smaller than the range of magnetic deviation maintaining metrological gain.
			
			{\centering
				\includegraphics[width=1\columnwidth]{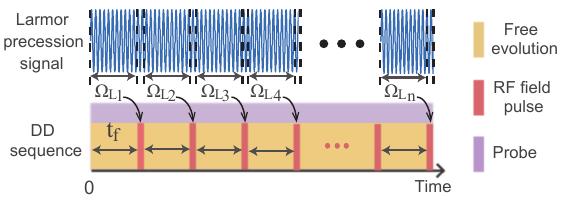}
				\caption{\textbf{The closed-loop scheme.} The blue line is the Larmor precession signal. The measurement process is implemented during the free evolution process lasting $t_{f}$. $\Omega_{Ln} (n=1,2,3...)$ are the measured real-time Larmor frequency. The frequency of the RF field pulse is adjusted to the Larmor frequency of the previous measurement.}
				\label{AP}}
			
			The single-shot measurement uncertainty is calculated by $1/(\sqrt{g}\gamma t_{f}\sqrt{2Nf})$. Here, $t_{f}=T/p$ is the free evolution time since the measurement process is implemented during the free evolution process as shown in Fig\,\ref{AP}. Based on the pulse length discussed in the Supplementary Note 5, $t_{f}$ could be 400\,$\mu$s. With atom number $N=1.42\times10^{10}$, the single-shot measurement uncertainty is 1\,pT.
			
			To evaluate the impact brought by the measurement uncertainty, we calculate the  metrological gain $g$ obtained from the initial SSS and the SSS after 200 DD cycles change at different magnetic field deviation ratios $\delta{B}/B_{0}$ within the range of [$-5\times 10^{-5}$, $5\times 10^{-5}$], corresponding to $\delta{B}$ in [-2.5\,nT, 2.5\,nT], as $B_0=50\,\mu$T (the Earth's magnetic field). The metrological gain is $g=(\Delta{B}_{\rm SQL}/\Delta{B}_{\rm SSS})^2=[\langle\hat{\widetilde{F}}_y\rangle/(\sqrt{2Nf}\Delta\hat{\widetilde{F}}_x)]^2$. In the frame rotating at the RF field frequency $\omega_{RF}$ which is set as the Larmor frequency $\omega_{RF}=\Omega_{L}=\gamma\,B_0$, the Hamiltonian becomes $\hat{\widetilde{H}}_{B}=\gamma\delta B\hat{\tilde{f}}_{z}-\hbar\omega\hat{\tilde{f}}_{z}^{2}$. We still utilize the formula of $\nu$ derived from $B_0$ and the DD sequence of Eq.\,(\ref{dd}) in the main text to generate and lock the SSS.
			The resulting metrological gains are shown in Fig.\,\ref{deltaB}. The generation process of the SSS contains only one RF field pulse to rotate the squeezing axis to the direction of the measurement quantity, while the DD process contains 200 DD cycles each has 10 pulses. Therefore, the metrological gain achieved by the locked SSS after 200 DD cycles is more easily to be affected by the magnetic field deviation ratios, exhibiting an oscillating feature, as shown in the red line of Fig.\,\ref{deltaB}. Even so, the metrological gain shows robustness in a wide range of $\delta{B}$ in [-50 pT, 50 pT], as shown in the insert of Fig.\,\ref{deltaB}. This range is much bigger than the single-shot measurement uncertainty of 1\,pT.
			
			{\centering
				\includegraphics[width=1\columnwidth]{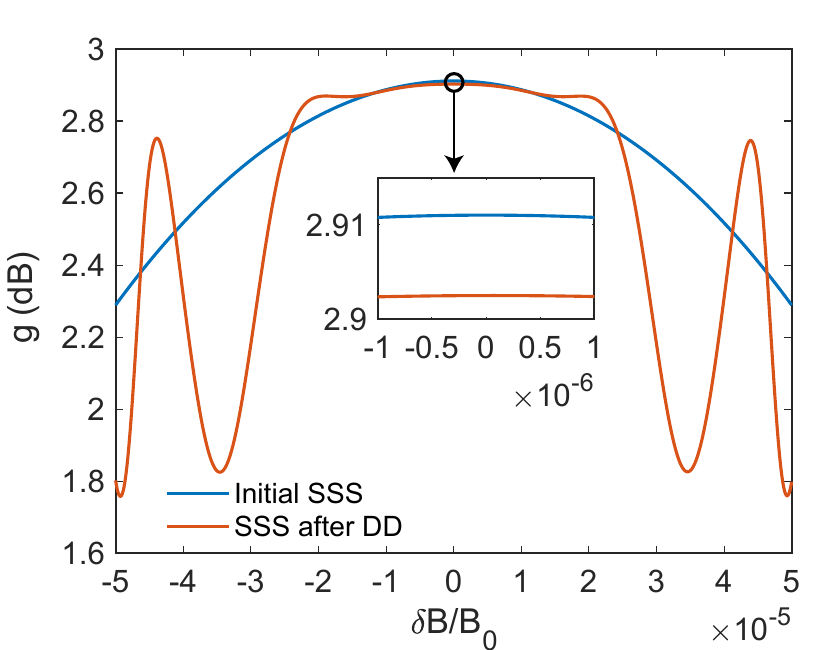}
				\caption{\textbf{The robustness analysis.} The metrological gain at different magnetic field deviation ratios $\delta{B}/B_0$. The blue line denotes the metrological gain achieved by the initially generated SSS. The red line denotes the metrological gain achieved by the locked SSS after 200 DD cycles. Insert shows the case of $\delta{B}$ in the range of [-50\,pT, 50\,pT] with $B_0=50\mu$T.}
				\label{deltaB}}

		\end{methods}
		
		\section*{References}
		\vspace{0.4in}
		\bibliographystyle{naturemag}

		\begin{addendum}
			\item[Data Availability] All data needed to evaluate the conclusions in the paper are present in the paper and
			the Supplementary Materials. Additional data related to this paper may be requested from the authors.
			\item 	We would like to thank D. Budker from Johannes Gutenberg University, Keye Zhang and Lu Zhou from East China Normal University for useful discussions. This work is supported by the Innovation Program for Quantum Science and Technology (2021ZD0303200); the National Natural Science Foundation of China (12234014, 11654005, 12204303, 11904227); the Fundamental Research Funds for the Central Universities; the Shanghai Municipal Science and Technology Major Project (2019SHZDZX01); the National Key Research and Development Program of China (Grant No. 2016YFA0302001); and the Fellowship of China Postdoctoral Science Foundation (Grant No. 2020TQ0193, 2021M702146, 2021M702150, 2021M702147, 2022T150413); W. Z. also acknowledges additional support from the Shanghai talent program.
			\item[Author Contributions] G. B. and W. Z. conceived the research. W. Z. supervised the whole project. G. B., P. Y., and J. C. carried out the calculations. G. B., P. Y., J. C., W. D., J. G. and W. Z. contributed to the theoretical analysis. G. B., P. Y., J. C. and W. Z. wrote the manuscript. All authors contributed to the discussion and review of the manuscript.
			\item[Competing Interests] The authors declare no competing financial or non-financial interests.
			\item[Correspondence] Correspondence and requests for materials
			should be addressed to G. B. (email: guzhibao@sjtu.edu.cn) and W. Z. (email: wpz@sjtu.edu.cn).
		\end{addendum}
		
	\end{multicols}
	\newpage
	\section*{\centering SUPPLEMENTARY NOTE 1: THE SQL FOR ATOM MAGNETOMETRY}
	In the condition discussed in the main text, the atomic system contains $N$ atoms each the magnetic quantum number $f$ and the measurement quantity is $\hat{\widetilde{F}}_x$. The SQL for atom magnetometry is defined as the sensitivity of the magnetic field measurement with CSS injection
	\begin{equation}
		\begin{aligned}
			\Delta B_{\rm SQL}=\frac{\Delta\hat{\widetilde{F}}_x}{|\frac{\partial\langle\hat{\widetilde{F}}_x\rangle}{\partial\theta_L}||\frac{\partial\theta_L}{\partial{B}}|}=\frac{\Delta\hat{\widetilde{F}}_x}{\gamma{T}_{2}\langle\hat{\widetilde{F}}_y\rangle}.
			\label{B1}
		\end{aligned}
	\end{equation}
	The atoms were initially polarized  in a way that their individual momenta were aligned along a common longitudinal direction, chosen as the $\hat{\tilde{y}}$ axis. This polarization results in a net magnetization $\vec{F}$ with magnitude $\vec{F}=\sqrt{F(F+1)}$, where $F=Nf$.
	The components of $\vec{F}$ in Cartesian coordinates are associated with noncommuting quantum operators, $\hat{\widetilde{F}}_x$, $\hat{\widetilde{F}}_y$ and $\hat{\widetilde{F}}_z$, which obey the Heisenberg uncertainty relation
	\begin{equation}
		\begin{aligned}
			\Delta\hat{\widetilde{F}}_z\Delta\hat{\widetilde{F}}_x\geq\frac{1}{2}|\hat{\widetilde{F}}_y|.
		\end{aligned}
	\end{equation}
	This inequality has the interpretation that an
	ensemble of measurements performed on either $\hat{\widetilde{F}}_x$ or $\hat{\widetilde{F}}_z$ will yield a distribution of random shot-to-shot outcomes.
	For a large magnetization, the $\hat{\widetilde{F}}_x$ measurement distribution is essentially Gaussian with mean $\langle\hat{\widetilde{F}}_x\rangle$ and variance $\Delta\hat{\widetilde{F}}_x^2=\langle\hat{\widetilde{F}}_x^2\rangle-\langle\hat{\widetilde{F}}_x\rangle^2$.
	For CSS,
	\begin{equation}
		\begin{aligned}
			\langle\hat{\widetilde{F}}_y\rangle=F,
			\label{B2}
		\end{aligned}
	\end{equation}
	and \begin{equation}
		\begin{aligned}
			\Delta\hat{\widetilde{F}}_x=\sqrt{\frac{F}{2}}.
			\label{B3}
		\end{aligned}
	\end{equation}
	Combining Eqs.(\ref{B1},\ref{B2},\ref{B3}), we have
	\begin{equation}
		\begin{aligned}
			\Delta B_{\rm SQL}=\frac{1}{\gamma{T}_{2}\sqrt{2F}}=\frac{1}{\gamma{T}_{2}\sqrt{2Nf}}.
		\end{aligned}
	\end{equation}

	\section*{\centering SUPPLEMENTARY NOTE 2: SPIN SQUEEZING OF SINGLE ATOM IN COUPLED BASIS}
	
	For $f=2$ system, we analyse the spin squeezing in the frame rotating at $\Omega_L$ by means of the unitary transformation operator $\hat{S}(t) = e^ {-i\hat{D}t}$,
	where
	\begin{equation}
		\begin{aligned}
			\hat{D}=\left( \begin{array}{ccccc}
				2\Omega_L & 0 & 0 & 0 & 0\\
				0 & \Omega_L & 0 & 0 & 0\\
				0 & 0 & 0 & 0 & 0\\
				0 & 0 & 0 & -\Omega_L & 0\\
				0 & 0 & 0 & 0 & -2\Omega_L\\
			\end{array}
			\right).
		\end{aligned}
	\end{equation}
	Then, the Hamiltonian of atom-magnetic field interaction in the rotating frame is
	\begin{equation}
		\hat{\widetilde{H}}_{B}=\hat{S}^{\dagger}(t) \hat{H}_{B} \hat{S}(t)-i\hbar\hat{S}^{\dagger}(t)\frac{\partial \hat{S}(t)}{\partial t}=-\hbar\omega \hat{\tilde{f}}_{z}^2.
	\end{equation}
	The angular momenta in the rotating frame are
	\begin{subequations}
		\begin{align}	
			\hat{\tilde{f}}_{x}=&\hat{S}^{\dagger}(t) \hat{f}_{x} \hat{S}(t)=\hat{f}_{x}\cos(\Omega_L t)+\hat{f}_{y}\sin(\Omega_L t),\\
			\hat{\tilde{f}}_{y}=&\hat{S}^{\dagger}(t) \hat{f}_{y} \hat{S}(t)=\hat{f}_{y}\cos(\Omega_L t)-\hat{f}_{x}\sin(\Omega_L t),\\
			\hat{\tilde{f}}_{z}=&\hat{S}^{\dagger}(t) \hat{f}_{z} \hat{S}(t)=\hat{f}_{z}.
		\end{align}
	\end{subequations}
	
	To see the spin squeezing effect, we analyse the operators $\hat{\tilde{f}}_{x}$ and $\hat{\tilde{f}}_{y}$ evolution in the Heisenberg picture. The unitary transformation of free evolution in the rotating frame is
	\begin{equation}
		\hat{\widetilde{U}}_{F}(t)=e ^{-i \hat{\widetilde{H}}_{B} t/\hbar}=e ^{i \omega t \hat{\tilde{f}}_{z}^2}.
		\label{E2}
	\end{equation}
	
	The raising and lowering operators are
	\begin{equation}
		\hat{\tilde{f}}_\pm(t)=\hat{\tilde{f}}_x(t)\pm i\hat{\tilde{f}}_y(t).
		\label{E3}
	\end{equation}
	evolve as
	\begin{equation}
		\begin{aligned}
			\hat{\tilde{f}}_{+}(t)= &\hat{\widetilde{U}}_{F}(t)^\dagger \hat{\tilde{f}}_+(0)\hat{\widetilde{U}}_{F}(t)\\
			= &e^{-i \omega t \hat{\tilde{f}}_{z}^2}\hat{\tilde{f}}_+(0)e^{i\omega t \hat{\tilde{f}}_{z}^2}\\
			=&\hat{\tilde{f}}_+(0)e^{-2i \omega t (\hat{\tilde{f}}_{z}+\frac{1}{2})},
		\end{aligned}
		\label{E4}
	\end{equation}
	\begin{equation}
		\hat{\tilde{f}}_-(t)=[\hat{\tilde{f}}_+(t)]^\dagger=e^{2i \omega t (\hat{\tilde{f}}_{z}+\frac{1}{2})}\hat{\tilde{f}}_-(0).
		\label{E5}
	\end{equation}
	
	According to Eq.\,(\ref{E3}), we have
	\begin{equation}
		\begin{aligned}
			\hat{\tilde{f}}_x(t)= &\frac{1}{2}[\hat{\tilde{f}}_+(t)+\hat{\tilde{f}}_-(t)]\\
			= &\frac{1}{2}[\hat{\tilde{f}}_+(0)e^{-2i \omega t (\hat{\tilde{f}}_{z}+\frac{1}{2})}+e^{2i \omega t (\hat{\tilde{f}}_{z}+\frac{1}{2})}\hat{\tilde{f}}_-(0)],
		\end{aligned}
		\label{E6}
	\end{equation}
	\begin{equation}
		\begin{aligned}
			\hat{\tilde{f}}_y(t)= &\frac{1}{2i}[\hat{\tilde{f}}_+(t)-\hat{\tilde{f}}_-(t)]\\
			= &\frac{1}{2}i[\hat{\tilde{f}}_+(0)e^{-2i \omega t (\hat{\tilde{f}}_{z}+\frac{1}{2})}-e^{2i \omega t (\hat{\tilde{f}}_{z}+\frac{1}{2})}\hat{f}_-(0)].
		\end{aligned}
		\label{E7}
	\end{equation}
	
	Next, we rotate $\hat{\tilde{f}}_x(t)$ and $ \hat{\tilde{f}}_{z}(t) $ around the -$ \hat{\tilde{y}} $ axis by $ \nu $, that is
	\begin{equation}
		\hat{\tilde{f}}_{x,v}(t)= e^{-i \nu \hat{\tilde{f}}_y(t)} \hat{\tilde{f}}_x(t) e^{i \nu \hat{\tilde{f}}_y(t)},
		\label{E8}
	\end{equation}
	\begin{equation}
		\hat{\tilde{f}}_{z,v}(t)= e^{-i \nu \hat{\tilde{f}}_y(t)} \hat{\tilde{f}}_{z}(t) e^{i \nu \hat{\tilde{f}}_y(t)}.
		\label{E9}
	\end{equation}
	As the initial state is along the $\hat{\tilde{y}}$ axis, the means and variances of the angular momentum are given by
	\begin{subequations}
		\begin{align}	
			\left\langle \hat{\tilde{f}}_{y}\right\rangle=& f\cos^{2f-1}(\omega t),\\
			\left\langle \hat{\tilde{f}}_{z,\nu}\right\rangle=&0,\\
			\left\langle \hat{\tilde{f}}_{x,\nu}\right\rangle=&0.
			\label{E10}
		\end{align}
	\end{subequations}
	\begin{subequations}
		\begin{align}	
			(\Delta \hat{\tilde{f}}_{y})^2=&\frac{f}{2} \{2f[1-\cos^{2(2f-1)}(\omega t)]-(f-\frac{1}{2})X\},\\
			(\Delta \hat{\tilde{f}}_{z,\nu})^2=&\frac{f}{2} \{1+\frac{1}{2}(f-\frac{1}{2})[X+\sqrt{(X^2+Y^2)}\cos(2\nu+2\delta)]\},\\
			(\Delta \hat{\tilde{f}}_{x,\nu})^2=&\frac{f}{2} \{1+\frac{1}{2}(f-\frac{1}{2}) [X-\sqrt{(X^2+Y^2)}\cos(2\nu+2\delta)]\},
			\label{E11}
		\end{align}
	\end{subequations}
	where $ X=1-\cos^{2f-2}(2\omega t) $, $ Y=4\sin(\omega t)\cos^{2f-2}(\omega t) $, and $ \nu = \pi/2-1/2\arctan(Y/X) $.
	
	\section*{\centering SUPPLEMENTARY NOTE 3: THE NUCLEAR-ELECTRONIC SPIN ENTANGLEMENT IN UNCOUPLED BASIS}
	
	In the main text and NOTE 2, we analysis the spin squeezing in the coupled basis $|f\,m_f\rangle$.
	To make the physics more straightforward, here we derive the  nuclear-electronic spin entanglement in uncoupled basis $|I\,J\,m_{I}\,m_{J}\rangle$.
	The Hamiltonian of atom in an external magnetic field along the $\hat{z}$ axis is \cite{Dline}:
	\begin{equation}
		\hat{H}_{\rm uc} =A\,\hat{i}\cdot\hat{j}+\frac{\mu_B}{\hbar}(g_J\hat{j}_z+g_I\hat{i}_z)B,
		\label{huc}
	\end{equation}
	where $A$ characterizes the strengths of the magnetic-dipole interaction, $\hat{i}$ is the nuclear spin, $\hat{j}$ is the electron spin, $g_J$ and $g_I$ are the electronic and nuclear Land${\rm\acute e}$ factors, respectively, $\mu_{B}$ is the Bohr magneton. For $^{87}\text{Rb}5^{2}\mathop S_{1/2}$, $I=3/2$, $J=1/2$, $g_J=2$, $g_I=-0.000995$ \cite{Dline}. In the Earth-field range, the interaction of the nuclear spin and the magnetic field in Eq.(\ref{huc}) is neglected.
	The Hamiltonian is then written as
	\begin{equation}
		\hat{H}_{\rm uc} \approx A\,\hat{i}\cdot\hat{j}+4\Omega_{L}\hat{j}_z,
	\end{equation}
	where $\Omega_{L}=\gamma\,B=(\mu_{B}B)/(2\hbar)$ is the Larmor frequency. In the uncoupled basis $|I\,J\,m_{I}\,m_{J}\rangle$, $\hat{i}\cdot\hat{j}$ and $\hat{j}_z$ are written as
	\begin{equation}
		\begin{aligned}
			&\hat{i}\cdot\hat{j}=
			\hat{i}_x\otimes\hat{j}_x+\hat{i}_y\otimes\hat{j}_y+\hat{i}_z\otimes\hat{j}_z\\&=\frac{1}{2}\left( \begin{array}{cccccccc}
				0 & 0 & 0 & 0 & 0 & \frac{\sqrt{3}}{2} & 0 & 0\\
				0 & 0 & 0 & 0 & \frac{\sqrt{3}}{2} & 0 & 1 & 0\\
				0 & 0 & 0 & 0 & 0 & 1 & 0 & \frac{\sqrt{3}}{2}\\
				0 & 0 & 0 & 0 & 0 & 0 & \frac{\sqrt{3}}{2} & 0\\
				0 & \frac{\sqrt{3}}{2} & 0 & 0 & 0 & 0 & 0 & 0\\
				\frac{\sqrt{3}}{2} & 0 & 1 & 0 & 0 & 0 & 0 & 0\\
				0 & 1 & 0 & \frac{\sqrt{3}}{2} & 0 & 0 & 0 & 0\\
				0 & 0 & \frac{\sqrt{3}}{2} & 0 & 0 & 0 & 0 & 0\\
			\end{array}
			\right)+\frac{1}{2}\left( \begin{array}{cccccccc}
				0 & 0 & 0 & 0 & 0 & -\frac{\sqrt{3}}{2} & 0 & 0\\
				0 & 0 & 0 & 0 & \frac{\sqrt{3}}{2} & 0 & -1 & 0\\
				0 & 0 & 0 & 0 & 0 & 1 & 0 & -\frac{\sqrt{3}}{2}\\
				0 & 0 & 0 & 0 & 0 & 0 & \frac{\sqrt{3}}{2} & 0\\
				0 & \frac{\sqrt{3}}{2} & 0 & 0 & 0 & 0 & 0 & 0\\
				-\frac{\sqrt{3}}{2} & 0 & 1 & 0 & 0 & 0 & 0 & 0\\
				0 & -1 & 0 & \frac{\sqrt{3}}{2} & 0 & 0 & 0 & 0\\
				0 & 0 & -\frac{\sqrt{3}}{2} & 0 & 0 & 0 & 0 & 0\\
			\end{array}
			\right)\\&+\frac{1}{2}\left( \begin{array}{cccccccc}
				\frac{3}{2} & 0 & 0 & 0 & 0 & 0 & 0 & 0\\
				0 & \frac{1}{2} & 0 & 0 & 0 & 0 & 0 & 0\\
				0 & 0 & -\frac{1}{2} & 0 & 0 & 0 & 0 & 0\\
				0 & 0 & 0 & -\frac{3}{2} & 0 & 0 & 0 & 0\\
				0 & 0 & 0 & 0 & -\frac{3}{2} & 0 & 0 & 0\\
				0 & 0 & 0 & 0 & 0 & -\frac{1}{2} & 0 & 0\\
				0 & 0 & 0 & 0 & 0 & 0 & \frac{1}{2} & 0\\
				0 & 0 & 0 & 0 & 0 & 0 & 0 & \frac{3}{2}\\
			\end{array}
			\right)=\frac{1}{2}\left( \begin{array}{cccccccc}
				\frac{3}{2} & 0 & 0 & 0 & 0 & 0 & 0 & 0\\
				0 & \frac{1}{2} & 0 & 0 & \sqrt{3} & 0 & 0 & 0\\
				0 & 0 & -\frac{1}{2} & 0 & 0 & 2 & 0 & 0\\
				0 & 0 & 0 & -\frac{3}{2} & 0 & 0 & \sqrt{3} & 0\\
				0 & \sqrt{3} & 0 & 0 & -\frac{3}{2} & 0 & 0 & 0\\
				0 & 0 & 2 & 0 & 0 & -\frac{1}{2} & 0 & 0\\
				0 & 0 & 0 & \sqrt{3} & 0 & 0 & \frac{1}{2} & 0\\
				0 & 0 & 0 & 0 & 0 & 0 & 0 & \frac{3}{2}\\
			\end{array}
			\right),
		\end{aligned}
	\end{equation}
	\begin{equation}
		\begin{aligned}
			\hat{j}_z=\frac{1}{2}\left( \begin{array}{cccccccc}
				1 & 0 & 0 & 0 & 0 & 0 & 0 & 0\\
				0 & 1 & 0 & 0 & 0 & 0 & 0 & 0\\
				0 & 0 & 1 & 0 & 0 & 0 & 0 & 0\\
				0 & 0 & 0 & 1 & 0 & 0 & 0 & 0\\
				0 & 0 & 0 & 0 & -1 & 0 & 0 & 0\\
				0 & 0 & 0 & 0 & 0 & -1 & 0 & 0\\
				0 & 0 & 0 & 0 & 0 & 0 & -1 & 0\\
				0 & 0 & 0 & 0 & 0 & 0 & 0 & -1\\
			\end{array}
			\right).
		\end{aligned}
	\end{equation}
	The Hamiltonian $\hat{H}_{\rm uc}$ in the uncoupled basis is then
	\begin{equation}
		\begin{aligned}
			\hat{H}_{\rm uc}=\left( \begin{array}{cccccccc}
				\frac{3A}{4}+2\Omega_L & 0 & 0 & 0 & 0 & 0 & 0 & 0\\
				0 & \frac{A}{4}+2\Omega_L & 0 & 0 & \frac{\sqrt{3}A}{2} & 0 & 0 & 0\\
				0 & 0 & -\frac{A}{4}+2\Omega_L & 0 & 0 & A & 0 & 0\\
				0 & 0 & 0 & -\frac{3A}{4}+2\Omega_L & 0 & 0 & \frac{\sqrt{3}A}{2} & 0\\
				0 & \frac{\sqrt{3}A}{2} & 0 & 0 & -\frac{3A}{4}-2\Omega_L & 0 & 0 & 0\\
				0 & 0 & A & 0 & 0 & -\frac{A}{4}-2\Omega_L & 0 & 0\\
				0 & 0 & 0 & \frac{\sqrt{3}A}{2} & 0 & 0 & \frac{A}{4}-2\Omega_L & 0\\
				0 & 0 & 0 & 0 & 0 & 0 & 0 & \frac{3A}{4}-2\Omega_L\\
			\end{array}
			\right).
		\end{aligned}
	\end{equation}
	
	The relationship between the coupled $|f\,m_f\rangle$ and uncoupled $|I\,J\,m_{I}\,m_{J}\rangle$ basis is $\phi_{\rm c}=U\phi_{\rm uc}$ and the matrixform relationship is
	\begin{equation}
		\begin{aligned}
			\left( \begin{array}{c}
				|2\,2\rangle \\
				|1\,1\rangle\\
				|1\,0\rangle\\
				|1\,-1\rangle \\
				|2\,1\rangle \\
				|2\,0\rangle\\
				|2\,-1\rangle\\
				|2\,-2\rangle \\
			\end{array}\right)=\left( \begin{array}{cccccccc}
				1 & 0 & 0 & 0 & 0 & 0 & 0 & 0\\
				0 & -\frac{1}{2} & 0 & 0 & \frac{\sqrt{3}}{2} & 0 & 0 & 0\\
				0 & 0 & -\frac{\sqrt{2}}{2} & 0 & 0 & \frac{\sqrt{2}}{2} & 0 & 0\\
				0 & 0 & 0 & -\frac{\sqrt{3}}{2} & 0 & 0 & \frac{1}{2} & 0\\
				0 & \frac{\sqrt{3}}{2} & 0 & 0 & \frac{1}{2} & 0 & 0 & 0\\
				0 & 0 & \frac{\sqrt{2}}{2} & 0 & 0 & \frac{\sqrt{2}}{2} & 0 & 0\\
				0 & 0 & 0 & \frac{1}{2} & 0 & 0 & \frac{\sqrt{3}}{2} & 0\\
				0 & 0 & 0 & 0 & 0 & 0 & 0 & 1\\
			\end{array}
			\right)\left( \begin{array}{c}
				|\frac{3}{2}\,\frac{1}{2}\,\frac{3}{2}\,\frac{1}{2}\rangle \\
				|\frac{3}{2}\,\frac{1}{2}\,\frac{1}{2}\,\frac{1}{2}\rangle\\
				|\frac{3}{2}\,\frac{1}{2}\,-\frac{1}{2}\,\frac{1}{2}\rangle\\
				|\frac{3}{2}\,\frac{1}{2}\,-\frac{3}{2}\,\frac{1}{2}\rangle \\
				|\frac{3}{2}\,\frac{1}{2}\,\frac{3}{2}\,-\frac{1}{2}\rangle \\
				|\frac{3}{2}\,\frac{1}{2}\,\frac{1}{2}\,-\frac{1}{2}\rangle\\
				|\frac{3}{2}\,\frac{1}{2}\,-\frac{1}{2}\,-\frac{1}{2}\rangle\\
				|\frac{3}{2}\,\frac{1}{2}\,-\frac{3}{2}\,-\frac{1}{2}\rangle \\
			\end{array}\right),
		\end{aligned}
	\end{equation}
	in which $U$ is the transformation matrix.
	The Hamiltonian $\hat{H}_{\rm c}$ in the coupled basis is
	\begin{equation}
		\begin{aligned}
			\hat{H}_{\rm c}=U\hat{H}_{\rm uc}U^{\dagger}=\left( \begin{array}{cccccccc}
				\frac{3A}{4}+2\Omega_L & 0 & 0 & 0 & 0 & 0 & 0 & 0\\
				0 & -\frac{5A}{4}-\Omega_L & 0 & 0 & -\sqrt{3}\Omega_L & 0 & 0 & 0\\
				0 & 0 & -\frac{5A}{4} & 0 & 0 & -2\Omega_L & 0 & 0\\
				0 & 0 & 0 & -\frac{5A}{4}+\Omega_L & 0 & 0 & -\sqrt{3}\Omega_L & 0\\
				0 & -\sqrt{3}\Omega_L & 0 & 0 & \frac{3A}{4}+\Omega_L & 0 & 0 & 0\\
				0 & 0 & -2\Omega_L & 0 & 0 & \frac{3A}{4} & 0 & 0\\
				0 & 0 & 0 & -\sqrt{3}\Omega_L & 0 & 0 & \frac{3A}{4}-\Omega_L & 0\\
				0 & 0 & 0 & 0 & 0 & 0 & 0 & \frac{3A}{4}-2\Omega_L\\
			\end{array}
			\right).
			\label{hc}
		\end{aligned}
	\end{equation}
	The hyperfine-structure energy splitting $\Delta = 2A$. The eigenvalues of this Hamiltonian $\hat{H}_{\rm c}$ are given by Breit-Rabi formula, leading to the Eq.(1) in the main text.
	
	In this work, the initial state is the CSS stretched along the $\hat{y}$ axis.
	For convenience, we rotate the quantization axis from the $\hat{z}$ axis to the $\hat{y}$ axis with the transformation matrix
	\begin{equation}
		\begin{aligned}
			U_{x}=\left( \begin{array}{cccccccc}
				\frac{1}{4} & -\frac{\sqrt{3}i}{4} & -\frac{\sqrt{3}}{4} & \frac{i}{4} & -\frac{i}{4} & -\frac{\sqrt{3}}{4} & \frac{\sqrt{3}i}{4} & \frac{1}{4}\\
				-\frac{\sqrt{3}i}{4} & -\frac{1}{8} & 0 & -\frac{\sqrt{3}}{8} & -\frac{3\sqrt{3}}{8} & 0 & -\frac{3}{8} & \frac{\sqrt{3}i}{4}\\
				-\frac{\sqrt{3}}{4} & 0 & \frac{1}{4} & 0 & 0 & -\frac{3}{4} & 0 & -\frac{\sqrt{3}}{4}\\
				\frac{i}{4} & -\frac{\sqrt{3}}{8} & 0 & \frac{5}{8} & -\frac{1}{8} & 0 & -\frac{3\sqrt{3}}{8} & -\frac{i}{4}\\
				-\frac{i}{4} & -\frac{3\sqrt{3}}{8} & 0 & -\frac{1}{8} & \frac{5}{8} & 0 & -\frac{\sqrt{3}}{8} & \frac{i}{4}\\
				-\frac{\sqrt{3}}{4} & 0 & -\frac{3}{4} & 0 & 0 & \frac{1}{4} & 0 & -\frac{\sqrt{3}}{4}\\
				\frac{\sqrt{3}i}{4} & -\frac{3}{8} & 0 & -\frac{3\sqrt{3}}{8} & -\frac{\sqrt{3}}{8} & 0 & -\frac{1}{8} & -\frac{\sqrt{3}i}{4}\\
				\frac{1}{4} & \frac{\sqrt{3}i}{4} & -\frac{\sqrt{3}}{4} & -\frac{i}{4} & \frac{i}{4} & -\frac{\sqrt{3}}{4} & -\frac{\sqrt{3}i}{4} & \frac{1}{4}\\
			\end{array}
			\right).
		\end{aligned}
	\end{equation}
	Then, the initial CSS in the uncouples basis is
	\begin{equation}
		\begin{aligned}
			\Phi_{\rm uc0}=\left(\begin{array}{c}
				1\\
				0\\
				0\\
				0\\
				0\\
				0\\
				0\\
				0\\
			\end{array}
			\right),
		\end{aligned}
	\end{equation}
	and the Hamiltonian is $\hat{H}_{\rm uc}^{y}=U_{x}\hat{H}_{\rm uc}U_{x}^{\dagger}$.
	\begin{figure}[htbp]
		\centering
		\includegraphics[width=0.5\columnwidth]{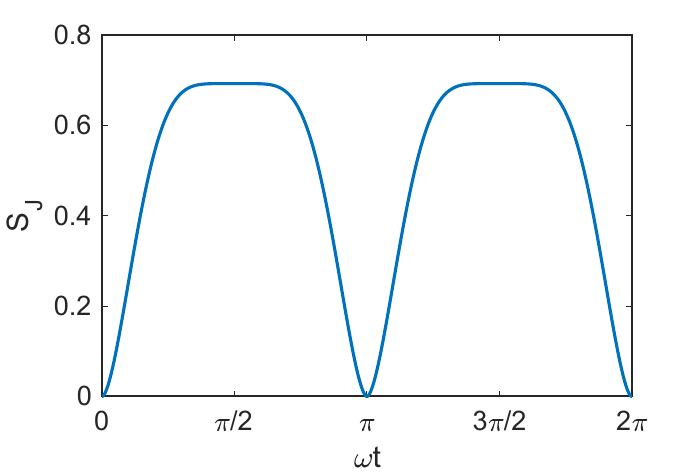} \caption{The entanglement entropy of the spin state evolving from the initial $\Phi_{\rm uc0}$ under the Hamiltonian $\hat{H}_{\rm uc}^{y}$ in the uncoupled basis. The values of the parameters are $\Omega_L=2\pi$, $\Omega_L/\omega=20000$ and $A=\Delta/2=\Omega_L^2/\omega/2$.}
		\label{Add}
	\end{figure}
	
	To characterize the  nuclear-electronic spin entanglement during the dynamic process, we calculate the entanglement entropy $S_{J}=-\text{Tr}[\rho_{J}\text{ln}\rho_{J}]$ of the spin state $\Phi_{\rm uc}(t)=e^{-\hat{H}^{y}_{\rm uc}t}\Phi_{\rm uc0}$ evolving from the initial $\Phi_{\rm uc0}$ under the Hamiltonian $\hat{H}_{\rm uc}$ \cite{Criterion}. Here, $\rho_{J}$ is the reduced density operator in the basis $|m_{J}=1/2\rangle,|m_{J}=-1/2\rangle$ yielded by tracing over the nuclear spin. The results are shown in Fig.\,\ref{Add}. The initial CSS $\Phi_{\rm uc0}$ at $\omega t=0$ is not entangled with a corresponding entanglement entropy of 0. At $\omega t=(2n+1)\pi/2$, where $n$ is an integer, the spin state evolves to
	\begin{equation}
		\begin{aligned}
			\Phi_{\rm uc}(t)=\frac{1}{2}\left(\begin{array}{c}
				1+i\\
				0\\
				0\\
				0\\
				0\\
				0\\
				0\\
				1-i\\
			\end{array}
			\right),
		\end{aligned}
	\end{equation}
	which is the maximum entanglement state with a corresponding entanglement entropy of $S_{J}=\text{ln}2$.

	\section*{\centering SUPPLEMENTARY NOTE 4: LOCKING PERFORMANCE WITH DIFFERENT TIME DEVIATIONS}
	
	\begin{figure}[htbp]
		\centering
		\includegraphics[width=0.5\columnwidth]{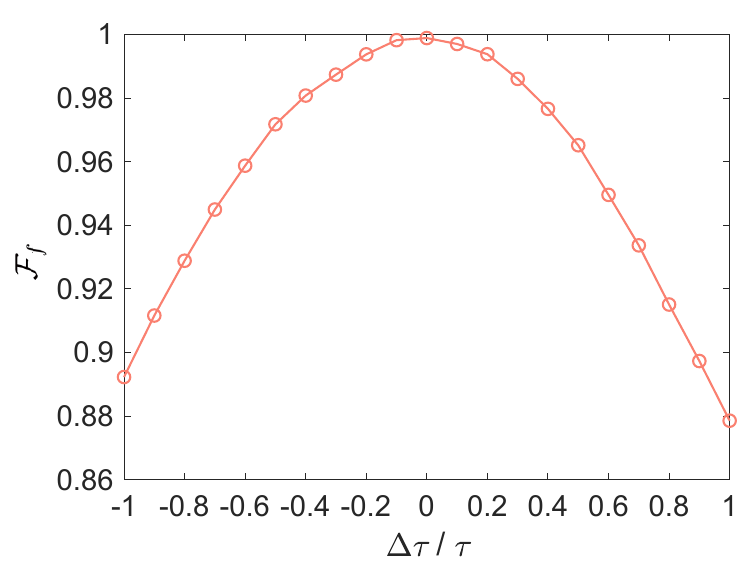} \caption{The fidelity with different time deviations.}
		\label{time}
	\end{figure}
	Since the time to lock the SSS might have deviations in the experiment, we study the locking performance of the quantum locking scheme with time deviations ratio $\Delta\tau/\tau$ ranging [-1,1]. The values of other parameters are $\Omega_{L}/\omega=20000$, $T\Omega_{L}=2\pi\times2000$, and $p=10$.  For each time deviation, the scheme outputs the corresponding optimal DD sequence. Figure \ref{time} shows the final fidelity at different time derivations, from which we can see that the fidelity keeps above 95\% in a wide range [-0.6,0.5]. It means that even if the optimization does not start at the optimal squeezing point, the quantum locking scheme can still lock the state to the optimal squeezing state.

	\section*{\centering SUPPLEMENTARY NOTE 5: THE VALUE OF THE PULSE LENGTH}
	
	According to the restrictive conditions for the parameters of DD sequence discussed in the main text: The pulse length $t_i$ should be considerably shorter than the free evolution time $t_{f}=T/p$ to satisfy the instaneous rotation condition and substantially longer than the Larmor period to satisfy the rotation wave approximation condition; The DD period $T$ should be substantially shorter than the revival period $2\pi/\omega$ to ensure that the SSS experiences enough DD cycles and keeps squeezed most of the time. As a result, we have $2\pi/\Omega_L\ll t_i$ and $t_i*p\ll T\ll 2\pi/\omega$. In $^{87}$Rb atom with the case of $p=10$, $\Omega_{L}=\gamma\,B_0$, $\omega=(\mu_{B}B_0)^2/(4\hbar^2\Delta)$, $\gamma=7*10^{9}$\,Hz/T, we have $1.4*10^{-10}({\rm T}\cdot {\rm s})/B_0\ll t_i\ll1.4*10^{-12}({\rm T}^2\cdot {\rm s})/B_0^2$. As $B_0=5*10^{-5}\,$T, $t_i\in$ [28 $\mu$s, 56 $\mu$s].

	\section*{\centering SUPPLEMENTARY NOTE 6: THE OPTIMIZATION PERFORMANCE FOR VARIOUS PULSE NUMBER p}
	
	\begin{figure}[htbp]
		\centering
		\includegraphics[width=0.5\columnwidth]{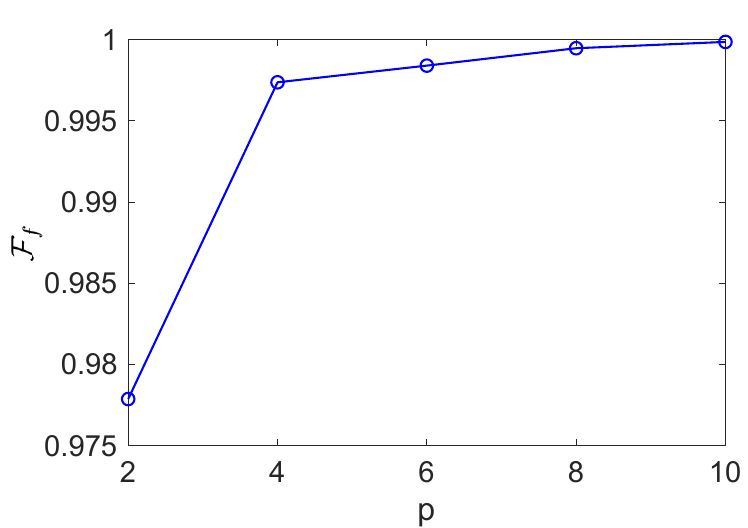} \caption{The fidelity at different values of the pulse number p.}
		\label{fidelity}
	\end{figure}
	To see how the number of the pulses in the DD sequence affect the optimization performance, we study the final fidelity at the pulse number $p = 2,4,6,8,10$. The optimal DD sequences are
	\begin{equation}
		\begin{aligned}
			\hat{\widetilde{U}}_{DD}(T)_{p=2}=\hat{\Pi}^{-\hat{\tilde{y}}}(\frac{\pi}{36})\hat{\widetilde{U}}_F(\frac{T}{2})
			\hat{\Pi}^{-\hat{\tilde{y}}}(\frac{\pi}{3})\hat{\widetilde{U}}_F(\frac{T}{2}),
		\end{aligned}
		\label{p2}
	\end{equation}
	\begin{equation}
		\begin{aligned}
			\hat{\widetilde{U}}_{DD}(T)_{p=4}& =\hat{\Pi}^{-\hat{\tilde{y}}}(\frac{17\pi}{36})\hat{\widetilde{U}}_F(\frac{T}{4})\hat{\Pi}^{-\hat{\tilde{y}}}(\frac{17\pi}{36}) \hat{\widetilde{U}}_F(\frac{T}{4})\\
			&\times\hat{\Pi}^{-\hat{\tilde{y}}}(\frac{7\pi}{36})\hat{\widetilde{U}}_F(\frac{T}{4})\hat{\Pi}^{-\hat{\tilde{y}}}(\frac{5\pi}{36})\hat{\widetilde{U}}_F(\frac{T}{4}),
		\end{aligned}
	\end{equation}
	\begin{equation}
		\begin{aligned}
			\hat{\widetilde{U}}_{DD}(T)_{p=6}& = \hat{\Pi}^{-\hat{\tilde{y}}}(\frac{5\pi}{36})\hat{\widetilde{U}}_F(\frac{T}{6})\hat{\Pi}^{-\hat{\tilde{y}}}(\frac{\pi}{2})\hat{\widetilde{U}}_F(\frac{T}{6})\\
			&\times\hat{\Pi}^{-\hat{\tilde{y}}}(\frac{\pi}{2})\hat{\widetilde{U}}_F(\frac{T}{6})\hat{\Pi}^{-\hat{\tilde{y}}}(\frac{\pi}{36}) \hat{\widetilde{U}}_F(\frac{T}{3})\\
			&\times\hat{\Pi}^{-\hat{\tilde{y}}}(\frac{7\pi}{36})\hat{\widetilde{U}}_F(\frac{T}{6}),
		\end{aligned}
	\end{equation}
	\begin{equation}
		\begin{aligned}
			\hat{\widetilde{U}}_{DD}(T)_{p=8}& =\hat{\Pi}^{-\hat{\tilde{y}}}(\frac{5\pi}{12})\hat{\widetilde{U}}_F(\frac{T}{8})\hat{\Pi}^{-\hat{\tilde{y}}}(\frac{\pi}{36})\hat{\widetilde{U}}_F(\frac{T}{8})\\
			&\times\hat{\Pi}^{-\hat{\tilde{y}}}(\frac{5\pi}{18})\hat{\widetilde{U}}_F(\frac{T}{8})\hat{\Pi}^{-\hat{\tilde{y}}}(\frac{13\pi}{36})\hat{\widetilde{U}}_F(\frac{T}{4})\\
			&\times\hat{\Pi}^{-\hat{\tilde{y}}}(\frac{\pi}{18}) \hat{\widetilde{U}}_F(\frac{T}{4})\hat{\Pi}^{-\hat{\tilde{y}}}(\frac{5\pi}{36})\hat{\widetilde{U}}_F(\frac{T}{8}),
		\end{aligned}
	\end{equation}
	\begin{equation}
		\begin{aligned}
			\hat{\widetilde{U}}_{DD}(T)_{p=10}& = \hat{\Pi}^{-\hat{\tilde{y}}}(\frac{\pi}{36})\hat{\widetilde{U}}_F(\frac{T}{10})\hat{\Pi}^{-\hat{\tilde{y}}}(\frac{4\pi}{9})\hat{\widetilde{U}}_F(\frac{T}{10})\\
			&\times\hat{\Pi}^{-\hat{\tilde{y}}}(\frac{\pi}{4})\hat{\widetilde{U}}_F(\frac{T}{10})\hat{\Pi}^{-\hat{\tilde{y}}}(\frac{4\pi}{9})\hat{\widetilde{U}}_F(\frac{T}{5})\\
			&\times\hat{\Pi}^{-\hat{\tilde{y}}}(\frac{\pi}{36})\hat{\widetilde{U}}_F(\frac{2T}{5})\hat{\Pi}^{-\hat{\tilde{y}}}(\frac{\pi}{6})\hat{\widetilde{U}}_F(\frac{T}{10}),
		\end{aligned}
	\end{equation}
	in which $T$ is set as 1/10 revival periods. The fidelity versus $p$ is shown in Fig.\,\ref{fidelity}.
	With the increase of $p$, the algorithm have better performance. The result for $p=10$ is good enough for the locking of SSS. So we choose $p=10$ for the DD sequence optimization.

	\section*{\centering SUPPLEMENTARY NOTE 7: THE EVOLUTION PATH OF THE OPTIMIZATION ALGORITHM}
	\begin{figure}[htbp]
		\centering
		\includegraphics[width=0.5\columnwidth]{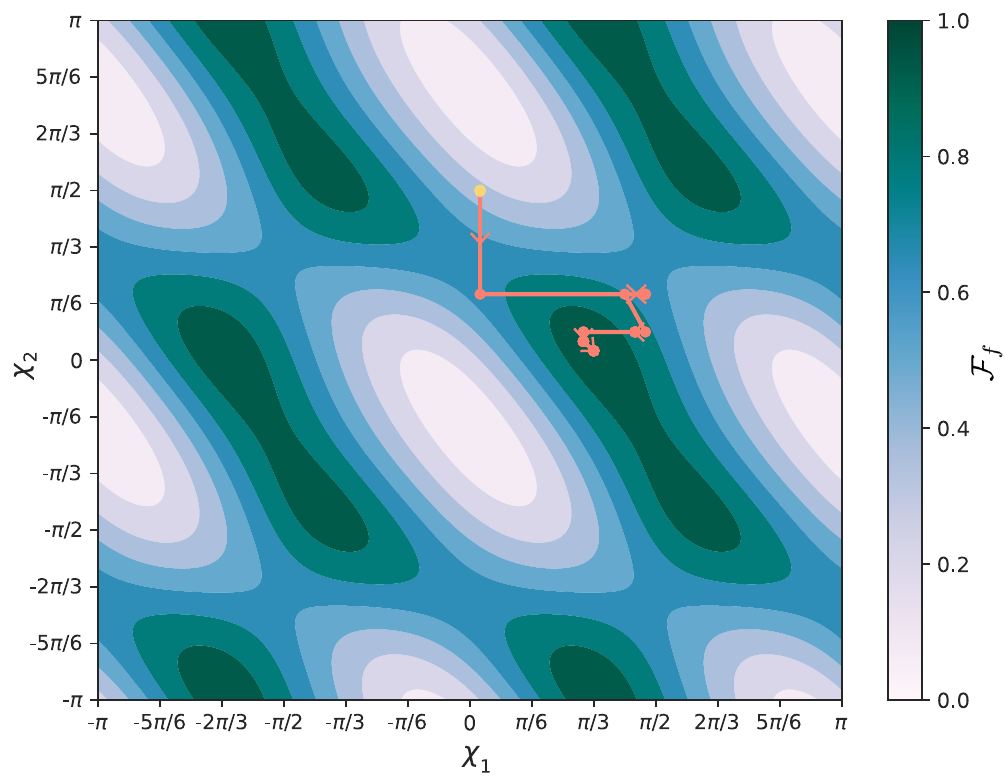} \caption{The landscape of the fidelity with possible rotation angles. The solid line denotes the evolution path of the optimal sequence. The yellow point is the random starting point.}
		\label{lanscape}
	\end{figure}
	To verify the predictive capability of the quantum locking technique, we apply the algorithm to a simple case: locking the SSS with $p=2$. Our goal is to determine the optimal rotation angles $\chi_1$ and $\chi_2$ in the DD sequence.
	The actual landscape is shown in Fig.\,\ref{lanscape} with the possible rotation angles. When the run number reaches 500, the result converges to the optimal DD sequence $\hat{\widetilde{U}}_{DD}(T)_{p=2}$ of Eq.\,(\ref{p2}).
	
	Through observing the evolution path of the optimal sequence in Fig.\,\ref{lanscape}, we could find that the initial random point tends to fall in an unknown area. Next it will explore the adjacent areas and gradually approach the ``valley'' of the landscape. All the other sequences in the population explore the landscape in the similar way. Each sequence adjusts the forward direction according to other sequences' exploration. As the exploration process goes on, the population retains the sequences who have explored the area close to the valley, and gradually converges to the optimal sequence. This approach can be seen as an exploitation of the known information. As mentioned above, this method is a global optimization which balances exploration and exploitation so that the algorithm can efficiently design a near-optimal pulse sequence to meet our requirements. For locking the SSS with $p=10$, the evolution process is the same.

	\section*{\centering SUPPLEMENTARY NOTE 8: THE PHYSICAL PRINCIPLE OF THE TWO-PULSE QUANTUM LOCKING SCHEME}
	
	In this section, we analyse the physical principle of the quantum locking technology for the two-pulse case. As mentioned in the main text, the physical principle of the quantum locking is taking the shortcut of the spin state evolution from $\Psi(t_1)$ to $\Psi(2\pi-t_1)$. Here, we show the detail that how the DD sequence optimized by the DE algorithm achieves the shortcut evolution.
	We analyse the density matrix evolution of the atomic spin in the $\hat{\tilde F}_y$ representation. The density matrix is written as
	\begin{equation}
		\begin{aligned}
			\tilde{\rho}=\left( \begin{array}{ccccc}
				\tilde{\rho}_{-2,-2} & \tilde{\rho}_{-2,-1} & \tilde{\rho}_{-2,0} & \tilde{\rho}_{-2,1} & \tilde{\rho}_{-2,2}\\
				\tilde{\rho}_{-1,-2} & \tilde{\rho}_{-1,-1} & \tilde{\rho}_{-1,0} & \tilde{\rho}_{-1,1} & \tilde{\rho}_{-1,2}\\
				\tilde{\rho}_{0,-2} & \tilde{\rho}_{0,-1} & \tilde{\rho}_{0,0} & \tilde{\rho}_{0,1} & \tilde{\rho}_{0,2}\\
				\tilde{\rho}_{1,-2} & \tilde{\rho}_{1,-1} & \tilde{\rho}_{1,0} & \tilde{\rho}_{1,1} & \tilde{\rho}_{1,2}\\
				\tilde{\rho}_{2,-2} & \tilde{\rho}_{2,-1} & \tilde{\rho}_{2,0} & \tilde{\rho}_{2,1} & \tilde{\rho}_{2,2}\\
			\end{array}
			\right),
		\end{aligned}
	\end{equation}
	where $\tilde{\rho}_{mn}$ is the density matrix element, $m$ and $n$ are magnetic quantum numbers. With the rotation pulse $\chi_{y}$ applied, a uniform expression for the transformation of density-matrix elements is
	\begin{equation}
		\tilde{\rho}_{mn}^{'}=e^{i(m-n)\chi_{y}}\tilde{\rho}_{mn}.
	\end{equation}
	
	\begin{figure}[htbp]
		\centering
		\includegraphics[width=0.5\columnwidth]{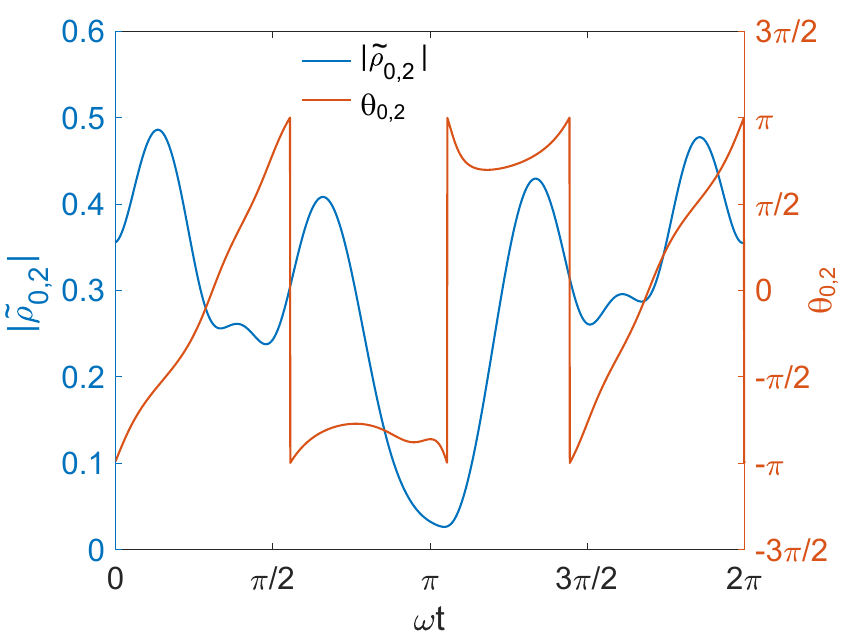} \caption{The evolution of the density matrix element $\tilde{\rho}_{0,2}$ with the NLZ effect. $|\tilde{\rho}_{0,2}|$ is the modulus and $\theta_{0,2}$ is the phase.}
		\label{rho}
	\end{figure}

		\begin{figure}[htpb]
		\includegraphics[width=1\columnwidth]{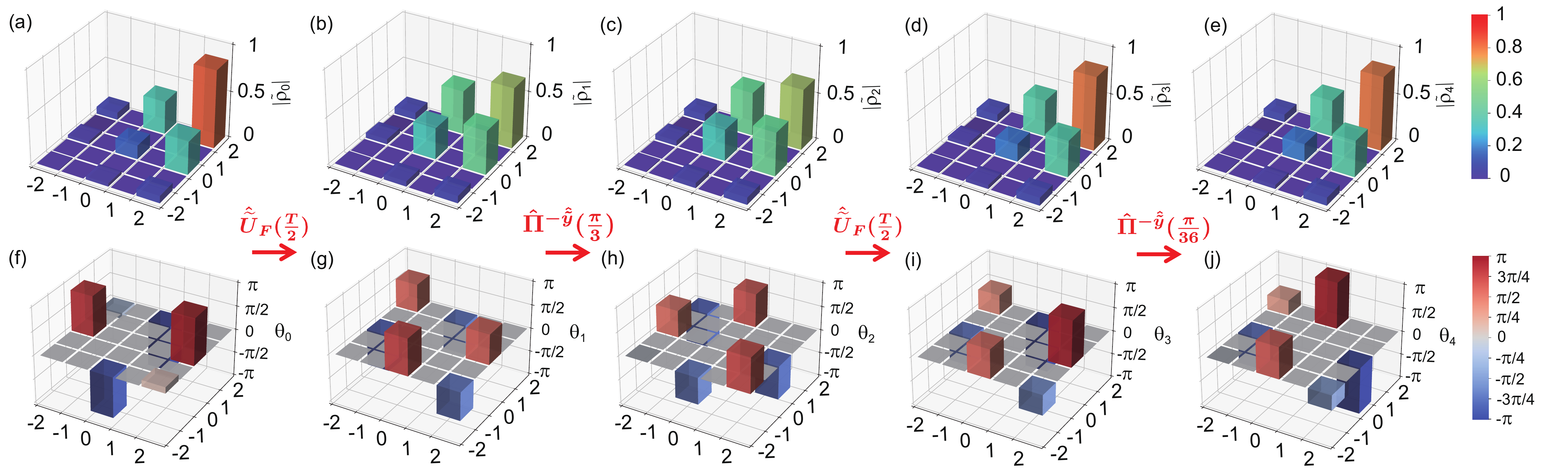} \caption{ The density matrix evolution under the DD sequence in the $\hat{\tilde F}_y$ representation. (a-e) are the modulus of the density matrix. (f-j) are the phase of the density matrix. $\tilde{\rho}_0$ is the SSS state. $\tilde{\rho}_4$ is the final state after one $\textrm{DD}$ sequence. The values of the parameters are $\Omega_{L}/\omega=20000$, $T\Omega_{L}=2\pi\times2000$, and $p=2$.}
		\label{density}
	\end{figure}
	The NLZ effect brings disturbances to both of the modulus and phase of the density matrix.
	Figure \ref{rho} shows the modulus $\lvert\tilde{\rho}_{0,2}\rvert$ and phase $\theta_{0, 2}$ of $\tilde{\rho}_{0,2}$ during the free evolution. We notice that for a short evolution time $t_{1}$, $\lvert\tilde{\rho}_{0,2}(t_1)\rvert=\lvert\tilde{\rho}_{0,2}(2\pi-t_1)\rvert$, $\theta_{0,2}(t_{1})=-\theta_{0,2}(2\pi-t_{1})$. By applying a $\chi_{y}$ pulse with $\chi_{y}=\theta_{0,2}(t_1)$, a transition from $\tilde{\rho}_{0,2}(t_1)$ to $\tilde{\rho}_{0,2}(2\pi-t_1)$ is achieved. This method suits for the other density matrix elements as well. The DE algorithm aims to balance the rotation angles for all the elements and achieves the shortcut from $\Psi(t_1)$ to $\Psi(2\pi-t_1)$ eventually.

	Taking the $p=2$ case as an example, Fig.\,\ref{density} shows the evolution of the density matrix with the calculated optimal DD sequence $\hat{\widetilde{U}}_{DD}(T)_{p=2}$.
	The phase of $\tilde{\rho}_{0,2}$ after the free evolution with evolution time $T/2$ is $\theta_{0,2}(T/2)=-\pi/3$.
	After the first rotation pulse $\hat{\Pi}^{-\hat{\tilde{y}}}(\frac{\pi}{3})$, the phase of $\tilde{\rho}_{0,2}$ becomes $\theta_{0,2}(2\pi-T/2)=\pi/3$. The second free evolution brings the $\tilde{\rho}_{0,2}$ back to the state of the initial SSS. Since the number of the rotation angles is set to 2, the DE algorithm only takes the main disturbed terms $\tilde{\rho}_{0,2}$ and  $\tilde{\rho}_{2,0}$ into consideration, giving the optimal solution.

	A successful quantum locking technique should be sufficient to bring all the disturbed terms to the initial state.
	For locking the CSS, only $\tilde{\rho}_{0,2}$ and $\tilde{\rho}_{2,0}$ need to be locked \cite{dd}. However for the SSS, more terms in the density matrix ($\tilde{\rho}_{0,2}$, $\tilde{\rho}_{2,0}$, $\tilde{\rho}_{2,-2}$, $\tilde{\rho}_{-2,2}$,) need to be locked. The single pulse can not make sure the phase of these terms reaches the condition of performing echo simultaneously. That's why a multi-pulse DD sequence is required for the locking of the SSS.
	
	\section*{References}
	\vspace{0.4in}
	\bibliographystyle{naturemag}

\end{document}